# Evidence for electron localisation in a moiré-of-moiré superlattice

Hangyeol Park[1,2,+], Junhyeok Oh[1,2,+], Rasoul Ghadimi[1,3], Chiranjit Mondal[1,3], Yungi Jeong[1,2], Won Beom Choi[1,2], Kenji Watanabe[4], Takashi Taniguchi[5], Bohm-Jung Yang[1,2,3]* and Joonho Jang[1,2]*

[1] Department of Physics and Astronomy, Seoul National University, Seoul 08826, Korea

[2] Institute of Applied Physics, Seoul National University, Seoul 08826, Korea

[3] Center for Theoretical Physics (CTP), Seoul National University, Seoul 08826, Korea

[4] Research Center for Electronic and Optical Materials, National Institute for Materials Science, 1-1 Namiki, Tsukuba 305-0044, Japan

[5] Research Center for Materials Nanoarchitectonics, National Institute for Materials Science, 1-1 Namiki, Tsukuba 305-0044, Japan

[+]These authors contributed equally to this work.
*correspondence to: joonho.jang@snu.ac.kr or bjyang@snu.ac.kr

**Abstract:** The localisation of electrons in a lattice potential is an intrinsically quantum-mechanical phenomenon and is often associated with remarkable physical properties of solids involving electron spins, electric polarisations and topological effects. In particular, even a small amount of distortion of the lattice potential can localise otherwise-delocalised quantum states in low-dimensional electron systems, dramatically influencing their thermodynamic properties and charge-transport behaviour. Study of such electron localisation induced by an aperiodic lattice potential remains exceptionally challenging in solid-state systems, since extrinsic disorders can trivially trap electrons in potential minima near disorders, obscuring the underlying quantum-mechanical origin of localisation phenomena. Van der Waals heterostructures can provide an alternative route for explorations of the phenomena via the emergence of superlattice potentials generated by rotating and stacking individual layers. Here, we report strong signatures of electron localisation in helical trilayer graphene, where the interplay of two moiré patterns gives rise to a moiré-of-moiré superlattice with distinct regions of moiré-periodic and moiré-aperiodic potentials. Remarkably, our measurements reveal the presence of double moiré-induced bands and high-order Brown–Zak oscillations, which are direct reflections of the periodic region with two constituent moiré patterns, and a superimposed anomalous hysteretic signal attributable to the aperiodic region. The data strongly suggest that electron wave functions are partially localised driven by the loss of a periodic lattice potential. Our work provides insight into the effects of spatially inhomogeneous lattice potentials on the low-dimensional electronic states and introduces a promising approach to control electron localisation for practical applications in solid-state devices.

**Introduction**

In the quantum theory of solids, the periodic atomic potential of a crystalline lattice supports '*delocalised*' extended Bloch waves for electrons to propagate coherently through an array of atoms. The theoretical framework lays the foundation for understanding electronic states in crystalline solids, and the distinction between metallic and insulating behaviour is fundamentally understood. However, the applicability of the Bloch theory can break down dramatically when the underlying periodicity is lost. Especially in systems whose dimensionality is equal or lower than two, the extended states can transform into localised ones in a subtle modification of the periodic lattice [1–5], highlighting that charge transport in low-dimensional systems is critically influenced by the loss of crystal translational symmetry. Further theoretical studies have shown explicitly that aperiodic potential modulations realisable in a few systems—such as quasicrystals or incommensurate moiré superlattices—indeed induce electron localisation [6–11]. Yet, despite these theoretical advances, electron localisation in low-dimensional aperiodic systems remains largely unexplored experimentally, owing in part to the difficulty of engineering aperiodic modulations in clean, high-quality materials. Investigating localisation in this critical dimensionality thus offers a rare opportunity to deepen our understanding of how quantum coherence and structural complexity govern the emergence of electronic phases in low-dimensional systems.

In this context, electronic systems based on van der Waals (vdW) heterostructures offer a promising platform for probing such phenomena. Their exceptional tunability—achieved through twisting and stacking multiple layers—enables the deliberate engineering of artificial structures, known as moiré superlattices. Potential modulations from this pattern modify the low-energy spectra of the electronic band structure [12–15] and have been shown to support various quantum many-body phases [16–25], wherein electron interactions, band topology, and their interplay play crucial roles. What makes twisted vdW heterostructures even more intriguing is that, when stacked with more than two layers, they host multiple moiré superlattices and facilitate experimental investigations into how emergent superstructures of moiré superlattices—via moiré relaxation—further modify the electronic spectra [26–36]. Notably, by tuning twist angles between each layer, interlayer commensurability can be favoured over intralayer distortion, leading to the formation of a remarkable configuration characterized by domains with commensurate moiré sites surrounded by aperiodic incommensurate domain boundaries [32–36]. These distinctive features render twisted vdW heterostructures particularly well-suited for systematically exploring electronic phases in aperiodic lattice potentials, overcoming limitations inherent to the constituent materials.

Among the series of twisted vdW systems, twisted three-layer graphene represents the simplest platform for achieving the moiré domain structure, as it naturally hosts two moiré superlattices generated by sequential twisting. In this work, we investigate helical trilayer graphene (hTG), a system consisting of three monolayer graphene sheets stacked sequentially with two equal twist angles. Strikingly, we observe hysteretic resistance responses across multiple devices, which we attribute to unusual transport phenomena along domain boundaries— regions with aperiodic lattice structures that emerge between adjacent moiré domains. Upon further investigation, we strongly suggest that electron localisation within these areas underlies the observed transport anomalies by modifying the local carrier density.

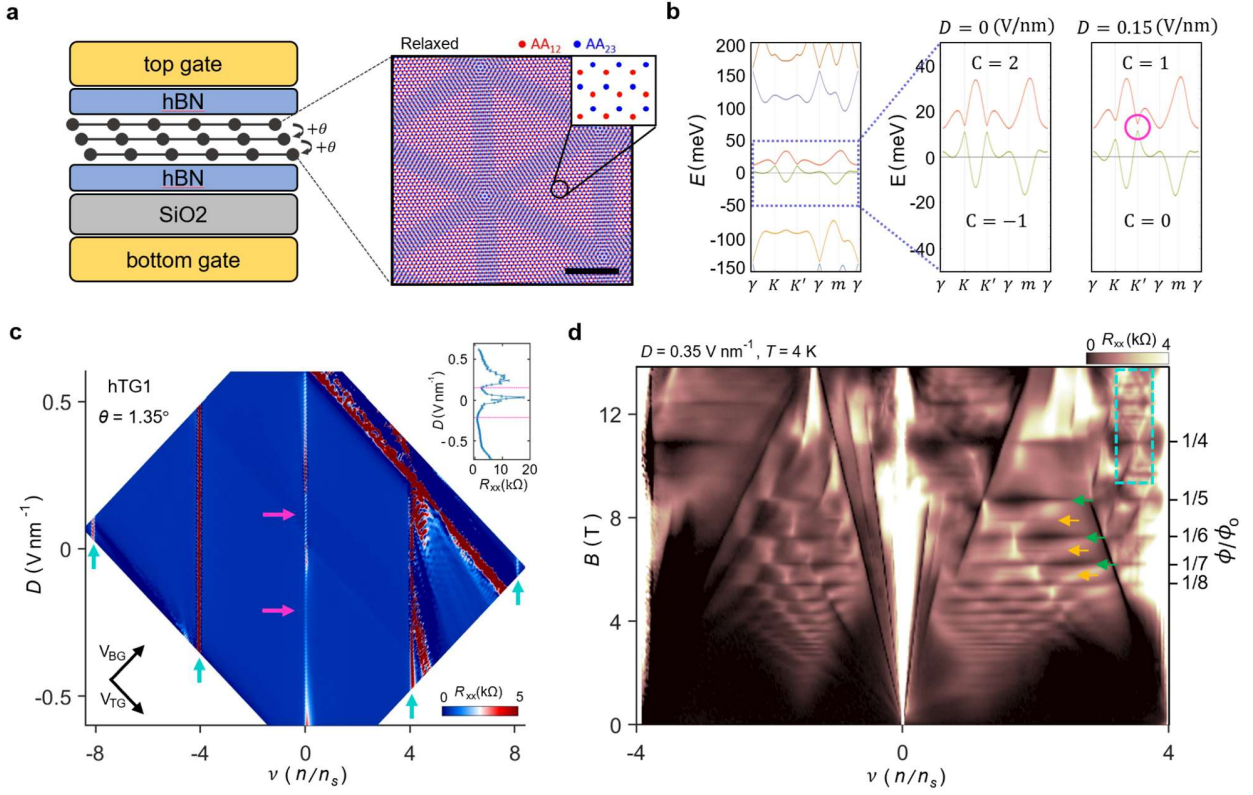

**Fig. 1: Lattice relaxation and characterization of hTG**
**a**, (Left) Schematic illustration of helical trilayer graphene (hTG), composed of three monolayer graphene sheets, each rotated by the same angle $\theta$ with identical helicity. (Right) Lattice configuration obtained from LAMMPS simulations of hTG at $\theta = 1.8°$, highlighting the effects of moiré relaxation (scale bar: 100 nm). Blue and red dots indicate the AA sites of the upper and lower moiré patterns, respectively. Atomic reconstruction leads to the formation of periodic domains separated by aperiodic domain boundaries, emphasised by shaded blue lines. Inset shows magnified view of a representative periodic domain. **b**, (Left) Electronic band structure at the K valley of a single domain in $\theta = 1.35°$ hTG at $D = 0$ V nm$^{-1}$, with each isolated band shown in a distinct colour. (Centre) Zoom-in of the region near the charge neutrality point (CNP), as indicated by the purple dashed box in the left panel. The two isolated bands carry Chern numbers of +2 and -1, respectively. (Right) At $D = 0.3$ V nm$^{-1}$, the band structure reflects a band inversion that has occurred at the location marked by the pink circle, with the associated Chern numbers having changed to +1 and 0, indicating a topological transition induced by the $D$ field. **c**, Longitudinal resistance map as a function of $n$ and $D$ at T = 2.5 K for the hTG1 device with a twist angle of 1.35°. The data show resistance peaks at the CNP and at moiré band insulators ($\nu = \pm 4$) as indicated by cyan arrows. At the CNP, two local resistance minima are highlighted by pink arrows. Inset shows line cut plot at the CNP, with two pink dotted lines indicating the positions marked by the pink arrows in the main panel. **d**, Landau fan diagram at $D = 0.35$ V nm$^{-1}$ and T = 4 K for the same device. The right y-axis represents the magnetic field normalized by the magnetic flux quantum. The green and orange arrows indicate the first- and second-order BZ oscillations, respectively, while higher-order BZ features appear within the cyan dotted box.

## Results

**Lattice relaxation and characterization of helical trilayer graphene**

We fabricated hBN-encapsulated hTG devices with dual gates (Methods), allowing control over both electronic density ($n$) and displacement field ($D$), with various twist angles ($\theta$) as illustrated in **Fig. 1a**. Without lattice relaxation, the two moiré patterns—formed between the top and middle graphene layers, and between the middle and bottom layers—are rotated by the same twist angle $\theta$. They are thus expected to generate a supermoiré pattern, so called moiré-of-moiré superstructure, whose characteristic length scale $\lambda_{sm} = \lambda_m/2\sin(\theta/2)$ is significantly larger than the original moiré period $\lambda_m = a/2\sin(\theta/2)$, where $a$ denotes the graphene lattice constant. However, it has been proposed that when the twist angle falls within a moderate range of approximately 1° to 3°, moiré lattice relaxation gives rise to triangular domains featuring a periodic arrangement of moiré sites (red and blue dots in the right panel of **Fig. 1a**), separated by domain boundaries where moiré sites appear at aperiodic positions [32,35]. The occurrence of this relaxation is further supported by our own simulation [30] (see also Methods) and was observed in a recent scanning probe measurement [33].

The periodicity of the domains allows their electronic band structure to be computed within the continuum approximation (see Methods). As an example, the calculated band structure for a single K valley at $D = 0$ V nm$^{-1}$ is presented in **Fig. 1b**. It shows that energy bands near the charge neutrality point (CNP) are well-separated by energy gaps reaching up to several tens of meV, and accommodates four electrons per moiré unit cell (**Extended Data Fig. 1**). Importantly, as the $D$ field increases (or decreases), the energy gap between the two isolated bands near the CNP gradually closes—leading to a band touching (at the point indicated by the pink circle in **Fig. 1b** for the increasing-$D$ case)—and then reopens. These gap-closing and reopening processes involve a change in the Chern numbers of the two bands, thereby corresponding to topological band inversions (see Supplementary Section II). Such theoretical predictions are well reflected in our transport measurements. **Fig. 1c** shows the measured four-probe longitudinal resistance map along the $n$–$D$ plane for the hTG1 device with a twist angle of 1.35°. The high-resistance vertical lines, indicated by cyan arrows, correspond to the moiré fillings at $\nu = 0, \pm 4$ and $\pm 8$, in agreement with the calculated band structure. A closer examination of the resistance at the CNP reveals two local minima as the $D$ field varies as indicated by pink arrows (see also inset for the line-cut plot), consistent with the presence of two band inversion points.

We also performed magnetotransport measurements. As shown in **Fig. 1d**, the magnetoresistance at $D = 0.35$ V nm$^{-1}$ and T = 4 K exhibits well-developed Landau fan features, along with Brown–Zak (BZ) oscillations persisting up to the fifth order (cyan dashed box and **Extended Data Fig. 2**). Clear Landau fan features reflect the high quality of the device, and the appearance of BZ oscillations is consistent with the underlying moiré periodicity inside the domains. In addition, key signatures—including high-resistance states appearing at moiré fillings that are multiples of four, and BZ oscillations—are reproducibly observed across multiple devices fabricated under comparable conditions, supporting that the phenomena reported in this work are intrinsic rather than device-specific artefacts (Supplementary Section I).

**Hysteretic transport in hTG**

While basic features in the measurement data of **Fig. 1** are well accounted for by the calculations, in **Fig. 2**, we observed a peculiar but distinct hysteretic behaviour, where the measured resistance values significantly changed depending on the sweep directions of the gate voltages. First, a resistance map measured as a function of the bottom gate voltage $V_{BG}$ (serving as the slow axis of 2D voltage sweeps, from -60 V to 60 V) and the top gate voltage $V_{TG}$ (serving as the fast axis, from -10 V to 10 V) (**Fig. 2a**; inset) is compared with another resistance map obtained using the same voltage ranges for both gates but with the bottom gate sweep direction reversed (**Fig. 2b**; inset). A direct comparison of these two resistance maps, especially in the regions enclosed by the cyan and orange dotted boxes, reveals a striking contrast in the resistance values of the insulating states.

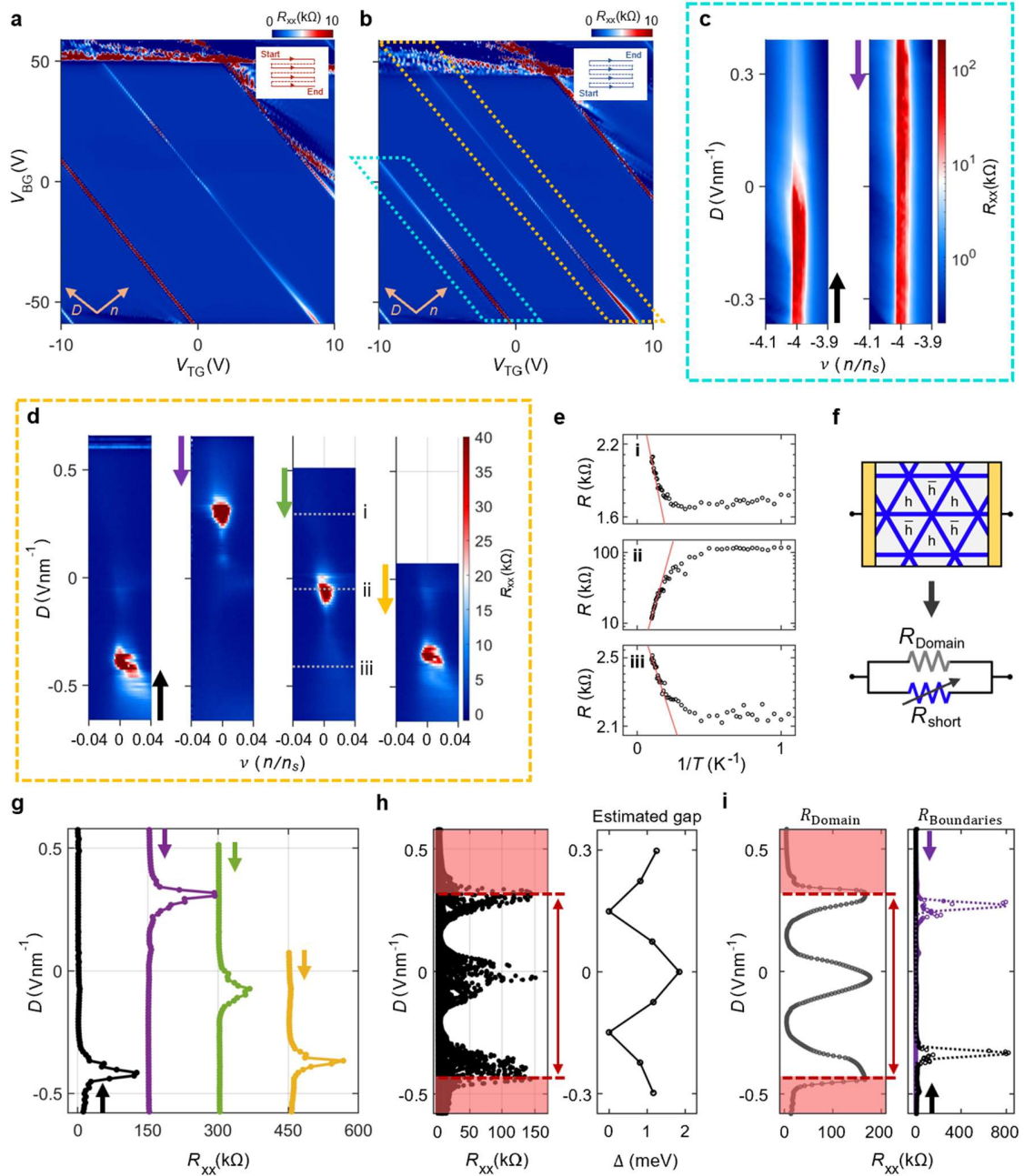

**Fig. 2: Hysteretic responses of total resistance**
**a–b**, Longitudinal resistance maps as a function of $V_{TG}$ and $V_{BG}$, measured during upward (a) and downward (b) sweeps of $V_{BG}$. In both maps, $V_{TG}$ was swept as the fast axis and $V_{BG}$ as the slow axis. High-resistance regions near $\nu = 0$ (cyan dotted box) and -4 (orange dotted box) exhibit clear hysteresis depending on the sweep direction. Insets show arrows indicating the direction of the gate sweeps. **c–d**, Resistance hysteresis near $\nu = -4$ (c) and $\nu = 0$ (d), measured over different ranges and directions of the $D$ field. Coloured solid arrows indicate the sweep directions and their respective starting point. **e**, Arrhenius plots of $R_{xx}$ near the CNP at $D = 0.3$ V nm$^{-1}$ (i), -0.1 V nm$^{-1}$ (ii) and -0.4 V nm$^{-1}$ (iii), all corresponding to the condition where the resistive packet resides near $D = 0$, as in the third panel of d. Among these, (ii) exhibits insulating behaviour, whereas (i) and (iii) remain metallic. **f**, Schematics of the toy model, represented as parallel-connected circuits, used to interpret the observed hysteresis. $R_{Domain}$ and $R_{Short}$ represent the resistance of the domains and domain boundaries, respectively. **g**, Line cuts of the resistance values as a function of $D$, extracted from the resistance maps at $n = 0$ in panel d. **h**, (Left) Superimposed resistance values at the CNP from multiple sweeps. The envelope within the region between the two red markers reflects the resistance of the domains. (Right) Simulated domain gap showing band inversion features as a function of $D$. The evolution of the simulated domain gap qualitatively agrees with the $D$ dependence of the domain resistance shown in the left panel. **i**, Extracted $R_{Domain}$ at the CNP (left) and that of the domain boundaries (Right). In the right panel, black and purple traces show representative upward and downward sweeps, respectively.

As shown in **Fig. 2c**, the difference between the upward- and downward-scans becomes more noticeable upon rescanning the area in $n$ and $D$ inside the cyan box of **Fig. 2b**, which corresponds to the insulating states near $\nu$ = -4. Notably, this insulating state is theoretically expected to host a substantial energy gap—on the order of tens of meV—regardless of the applied $D$ field (see Supplementary Section II), which would typically result in a resistance significantly larger than the quantum resistance, $h/e^2$ [37,38]. Nevertheless, the upward sweep in the left panel of **Fig. 2c** surprisingly reveals a pronounced reduction of resistance in the $D > 0$ V nm$^{-1}$ region in contrast to the same region in the downward sweep shown on the right. We interpret this low-resistance feature as a signature of anomalous metallic transport, considering the presence of a band gap at $\nu$ = -4 for the commensurate domains. The metallic behaviour is further corroborated by the observed temperature dependence of resistance, which decreases as the temperature is lowered (**Extended Data Fig. 3**).

A qualitatively similar resistance response is also observed near the CNP, corresponding to the region enclosed by the orange dotted box in **Fig. 2b**. This region is examined in greater detail through a series of measurements presented in **Fig. 2d**, taken under various sweep ranges and directions. Each sweep apparently reveals a small region of high resistance values—referred to here as a resistive packet—whose position shifts depending on gate sweep conditions. Based on the calculation in **Fig. 1b**, the domains have an energy gap at the CNP in most of the $D$ field values (except at discrete band inversion points); if the entire device were of a single domain, the resistance value at the CNP would exhibit insulating behaviour. Unexpectedly, however, the resistance at the CNP is fairly low except in the case that the $D$ field is tuned near the value of the resistive packet. The follow-up temperature dependence measurements confirm that the system is insulating only when it is tuned to be at the resistive packet (**Fig. 2e**).

This unusual transport response cannot be attributed solely to the electron transport in the commensurate domains of the hTG, which supposedly have band gaps at $\nu$ = -4 and 0 (CNP). Instead, the data point to the existence of an additional conductive channel that operates effectively in parallel to the conduction through the domains, thereby reducing the overall resistance, as illustrated schematically in the lower panel of **Fig. 2f**. This parallel channel appears to exhibit a hysteretic response to the gate voltage sweep. At $\nu$ = -4 or 0, while the commensurate domains are nearly insulating with band gaps, as the phenomenology of $R_{xx}$ map in **Fig. 1c** strongly implies, the parallel conducting channel can lower the device resistance, effectively masking the insulating response of the domains. On the other hand, when this parallel channel becomes highly insulating, the total device resistance is close to that of the domains; i.e., that expected from the calculated band structure.

If this interpretation is correct, one should be able to isolate (extract) the 'true' conduction of the domains from the hysteretic parallel conduction by tuning the condition so that the parallel conduction becomes highly resistive. **Fig. 2g** shows the device resistance measured at the CNP, by varying resistive packets' positions. Our analysis shows that the parallel path indeed becomes highly-resistive near the packet, and therefore the total device resistance in that regime is limited by the domain resistance (Supplementary Section III). Compiling the resistance data by varying the packet's positions, in the left panel of **Fig. 2h**, it becomes apparent that the resistance is certainly limited by a $D$ field dependent 'envelope' profile, which we attribute to the resistance contribution of the domains.

Certainly, this envelope is well explained by the evolution of the band gap at the CNP predicted by our calculations on the commensurate domains—especially its closing and reopening as a function of the $D$ field as shown in the right panel of **Fig. 2h**— thereby supporting our interpretation.

As a result, we can separately extract the resistance of the domains at the CNP and that of the parallel conduction, as shown in **Fig. 2i** (see also Supplementary Section III). The left panel shows the domain resistance, while the right panel presents the resistances of the parallel conduction. The latter is plotted as two representative traces in black and purple, corresponding to the upward and downward sweeps, respectively. The profiles exhibit a local maximum at the packet position and remain low elsewhere, yielding a Gaussian-like dependence on the $D$ field.

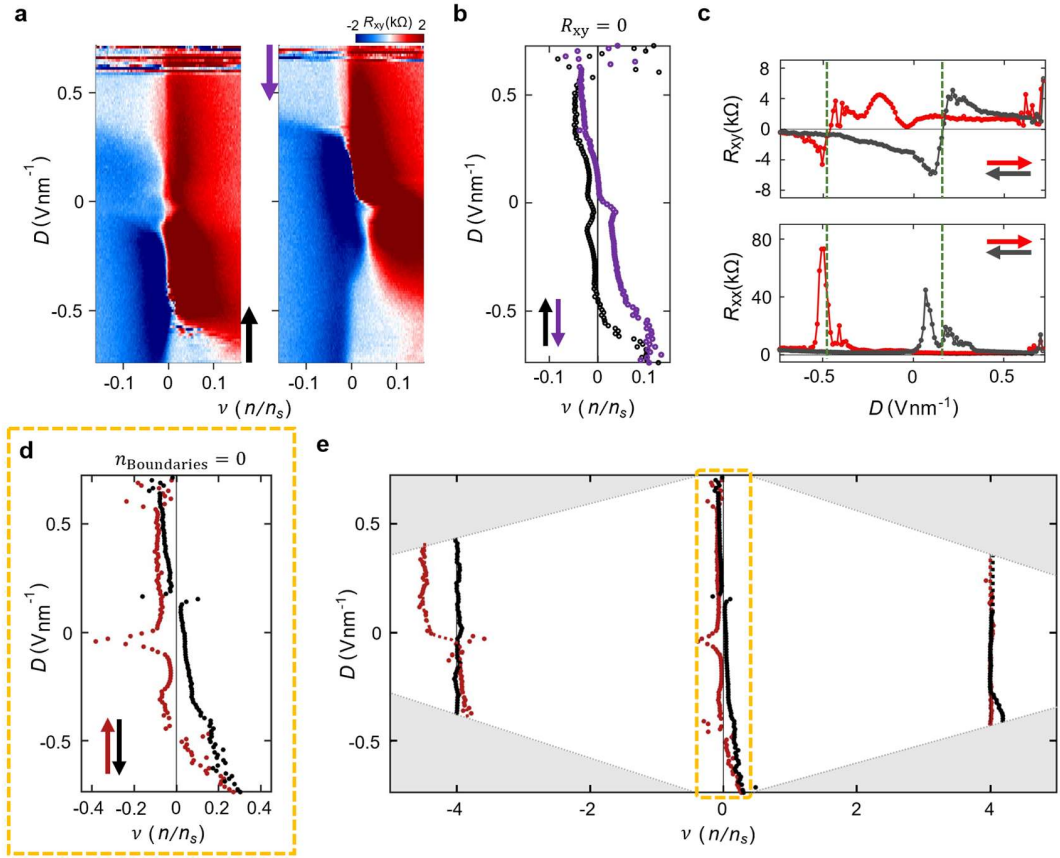

**Fig. 3: Hall measurements revealing hysteretic charge filling in domain boundaries**
**a**, Antisymmetrized Hall resistance, $R_{xy} = (R_{xy}(B_+)-R_{xy}(B_-))/2$, as a function of $\nu$ and $D$ measured at $B_\pm = \pm 0.2$ T and T = 4 K for the hTG1 device during upward (black) and downward (purple) sweeps. **b**, $R_{xy} = 0$ trajectories extracted from panel a for the upward (black) and downward (purple) sweeps, highlighting a hysteretic shift between the two sweep directions. Both trajectories deviate from the vertical grey line, which marks the CNP of the domains. **c**, Line cuts of $R_{xy}$ (top) and $R_{xx}$ (bottom) at the CNP for each sweep. The green dotted lines mark the positions where $R_{xy} = 0$, coinciding with peaks in $R_{xx}$. **d**, Estimated CNPs of the domain boundaries during upward and downward sweeps, indicated by red and black dots. **e**, Collective plots of the extracted CNPs of the domain boundaries at $\nu$ = -4, 0 and 4, plotted as a function of $n$ and $D$.

It is worth mentioning that the hysteresis is consistently reproduced across multiple high-quality hTG devices, making it unlikely that the shorting path arises from extrinsic artefacts. Instead, given the known relaxation pattern of hTG, it is natural to conclude that the domain boundaries in relaxed hTG are the origin of the hysteretic parallel conduction path. As depicted in **Fig. 2f**, the domain boundaries naturally form an interconnected network of conductive channels that bypass the gapped domains, effectively acting as a parallel pathway for transport. Notably, no comparable hysteresis is observed in our other moiré graphene systems, fabricated using the same process but lacking such domain boundaries—such as alternating twisted trilayer graphene (aTG), where the stacking configuration precludes their formation (Supplementary Section I). Together, the structural features of our system and experimental observations provide compelling evidence that the anomalous conduction originates from the domain boundaries.

**Hall measurements reveal hysteretic charge filling in domain boundaries**

To further explore the nature of the hysteretic signal, we performed Hall measurements, which offer useful insight into the charge carrier density and its variation with gate sweep conditions. In **Fig. 3a**, the antisymmetrized Hall resistance maps for upward and downward gate voltage sweeps are shown. As in the case of the longitudinal resistance data, the Hall resistance exhibits clear hysteresis, and it is more clearly highlighted in **Fig. 3b**, which shows the loci of points where the measured Hall resistance vanishes ($R_{xy} = 0$) for each sweep. Not only do the $R_{xy} = 0$ contours differ apparently between the two sweeps, but they also deviate from the $\nu = 0$ line—which corresponds to the charge neutrality condition of the domains—except for a couple of intersections. In contrast, the resistive packets consistently appear along the $\nu = 0$ line, regardless of the sweep conditions (see Supplementary Section III).

We also examined line-cuts of both the Hall ($R_{xy}$) and longitudinal ($R_{xx}$) resistances for each sweep along the $\nu = 0$ trajectory, as shown in the top and bottom panels of **Fig. 3c**. Notably, $R_{xy}$ vanishes only at the position where $R_{xx}$ reaches a maximum—the location of the resistive packet—while being non-zero at other points along the trajectory. Since the non-zero value of $R_{xy}$ implies the presence of net mobile carriers, these striking observations suggest that the system is not charge neutral along the $\nu = 0$ line, except for the $D$ values where the resistive packet appears. Given that the (commensurate) domains themselves are charge neutral along the $\nu = 0$ line, these residual carriers must be located in the areas between domains, i.e., in the domain boundaries. We therefore conclude that the resistive packet marks the special condition where both the domains and the domain boundaries are charge neutral, whereas in surrounding regions at the CNP, the domains remain neutral but their boundaries are doped.

The local carrier density of the domain boundaries ($n_{\text{Boundaries}}$) can be estimated from the measured Hall resistance. Because the domains are insulating along the $\nu = 0$ trajectory, electrical conduction is largely confined to the boundaries, making the measured Hall response a direct probe of their carrier content. Indeed, finite-element simulations confirm that the carrier density scales approximately inversely with the Hall resistance, following $n = B/(eR_{xy})$, regardless of the macroscopic network geometry (see Supplementary Section IV). Using the simulation, we extracted the carrier density of the domain boundaries at $\nu = 0$ for each $D$ field value and identified the gate voltage conditions at which they become charge neutral (see Supplementary Section III). The estimated neutrality points, obtained from upward and downward sweeps, are plotted in **Fig. 3d** using red and blue markers, respectively.

These neutrality trajectories trace out a closed loop that evolves with the applied $D$ field. We further carried out analogous measurements near $\nu = \pm 4$, and the results for all three filling factors are summarised in **Fig. 3e** (see Supplementary Section III). While the shapes of these trajectories differ between $\nu$ = -4, 0 and 4, they all exhibit hysteresis. Taken together, these results indicate that the sweep-induced evolution of the charge neutrality condition of the domain boundaries underlies the anomalous transport behaviours observed in hTG.

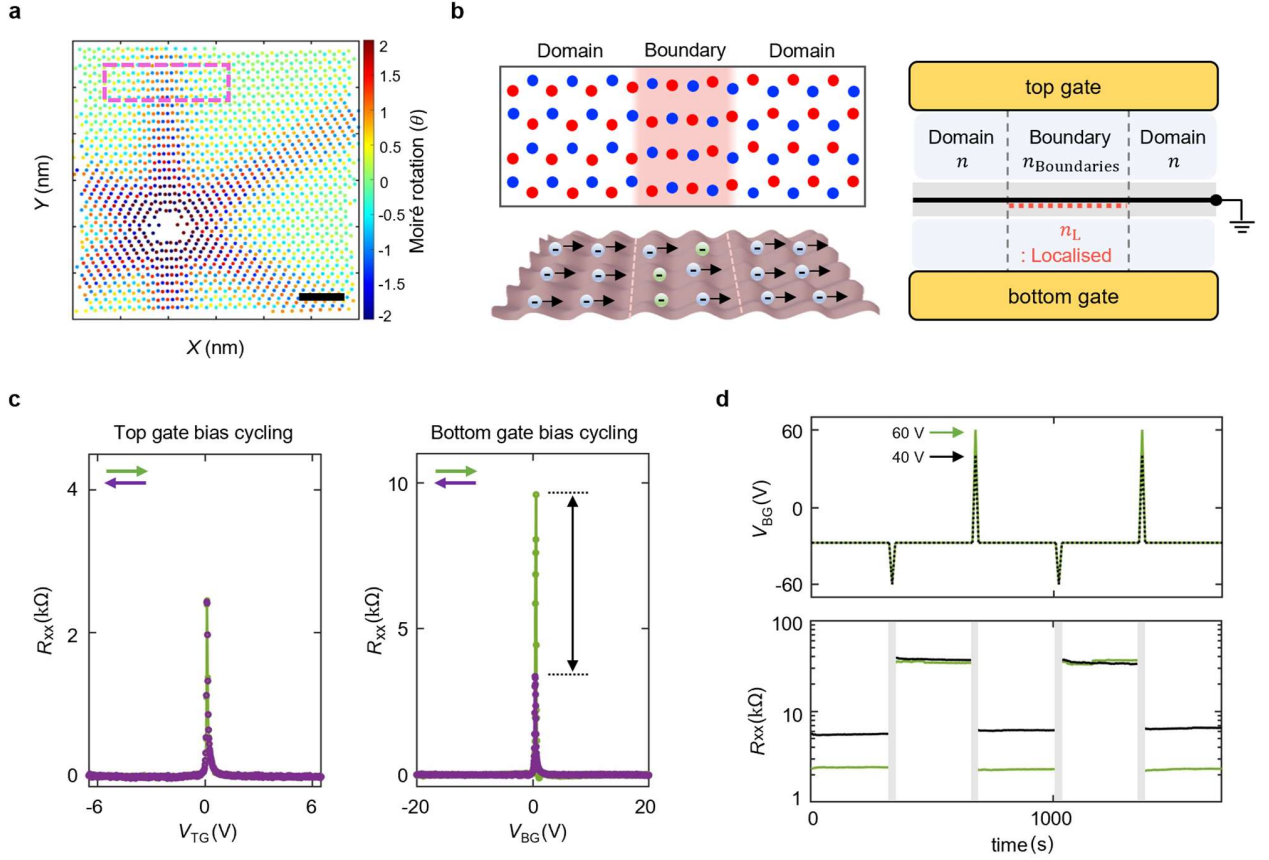

**Fig. 4: Electron localisation as a possible origin of the anomalous transport of hTG**
**a**, Lattice simulation coloured by local moiré rotation angle (scale bar represents 50 nm). **b**, (Left) Schematic illustration of the moiré lattice together with its corresponding electron distribution. The domains host only mobile carriers, whereas both mobile and localised carriers coexist along the domain boundaries. (Right) Side-view schematic illustrating the coexistence of localised states and mobile carriers along the domain boundaries. The charge density of localised states is denoted as $n_L$, while the domains and boundaries host distinct charge densities, $n$ and $n_{Boundaries}$, which typically differ owing to the presence of the localised states. **c**, Resistance measurements at 50 mK during cyclic sweeps of $V_{TG}$ (left, from -10 V to 10 V) and $V_{BG}$ (right, from -60 V to 60 V), with the opposite gate held fixed at 0 V. Hysteresis is observed only during $V_{BG}$ sweeps, indicating a gate-asymmetric response. **d**, $R_{xx}$ responses (bottom) to a series of $V_{BG}$ excitations (top) with varying voltage pulse amplitudes. As examples, the black and green traces show the time evolution of $R_{xx}$ in response to gate voltage excitation reaching 40 V and 60 V, respectively, each driving the system into distinct stable resistance states.

**Localised states as a possible origin of anomalous transport**

A key question emerges from our observation of gate-sweep-dependent charge filling: what is the origin of this behaviour? In hTG, the global inversion symmetry is inherently broken due to its helical stacking configuration in contrast to its cousin, alternating trilayer graphene (aTG). This structural feature may evoke interpretations of the observed behaviour in terms of ferroelectric switching, which is known to produce hysteresis through charge redistribution [39–41] or relative interlayer sliding [42–45]. However, these ferroelectric systems are not known to exhibit changes in the number of mobile carriers; they merely generate polarisation through ion displacement or sliding-induced structural rearrangement, which in turn modifies the effective $D$ field without altering the carrier density itself. This stands in contrast to our measurements, which reveal a pronounced modulation—on the order of $10^{11}$ cm$^{-2}$—in the number of mobile carriers participating in transport, depending on the sweep history. This clear distinction rules out ferroelectricity as a plausible explanation for the observed hysteresis.

Instead, we argue that this hysteretic charge density offset in $n_{\text{Boundaries}}$ (see **Fig. 3d**) is more naturally explained by electron localisation [46–55] arising from deformation of the periodic moiré potential at the domain boundaries. The lattice relaxation simulation (**Fig. 4a**) of the twisting angle of 1.4° reveals that, even though the hTG is stacked to have globally-twisted two rigid moiré lattices, the hTG readily relaxes to form periodic commensurate order in the domains while leaving the effect of twisting to be concentrated in the areas between the domains—i.e., domain boundaries—with strongly distorted moiré lattices. The previous theoretical works [6–11] showed that the moiré-incommensurability induces partial electron localisation; therefore, a portion of the electronic states in the domain boundaries of the hTG device are expected to be localised while the electronic states are delocalised inside domains. This theoretical framework supports our picture that the hysteretic charge density originates from the boundary area between domains.

Our observation appears closely related to the recent experimental works in the literature [50–55] that the hysteretic responses due to the localised states in the presence of moiré potentials were reported. In the works, localised states do not directly participate in conduction, but they can still affect the conducting channel indirectly; i.e., occupation of the localised states ($n_L$ in **Fig. 4b**) alters the local electrostatic environment, influencing mobile carriers involved in transport. What distinguishes our observation from the previous reports is that the effect of the localised states is limited to part of the sample, which is the moiré-incommensurate domain boundaries. The aperiodic boundary regions are envisioned to host both mobile carriers and localised electrons, with the latter representing the states that remain decoupled from the transport channel, as depicted in **Fig. 4b**.

Hence, under the action of globally uniform gates, there is a carrier density difference between the domains and their boundaries. This marked spatial non-uniformity cannot be explained by extrinsic mechanisms—such as defect trapping in hBN [46], interface traps [47,50,55], or dielectric polarisations [56]—which would affect the system uniformly. The observed spatial confinement instead points to an intrinsic origin in the structure of the moiré-of-moiré lattice. Furthermore, two additional experimental features further support the existence of localised states in our system: We observe a pronounced gate asymmetry in the hysteretic response, as shown in **Fig. 4c** (see also **Extended Data Figs. 4** and **5**), and stable multiple resistance states as shown in **Fig. 4d**, both of which are previously reported in low-dimensional materials hosting localised states [50–55].

**Discussion**

Our observations highlight lattice-relaxed hTG as a distinctive platform in which electron localisation emerges deterministically at the domain boundaries. It is well established that the long-length scale moiré patterns in twisted graphenes strongly renormalize low-lying energy states near the CNP into moiré-induced bands with reduced bandwidth [12]. In the same vein, within our hTG samples, the two moiré patterns interact with each other to form commensurate domains and incommensurate boundaries, leading to the partial localisation of the states in the moiré induced bands. In our view, this interplay of two moiré lattices account for the anomalous hysteresis reported in other hBN-aligned graphene heterostructures [50,54,55], where localised states have been inferred yet their microscopic origins remain unclear. The electron localisation—possibly arising from an aperiodic (or quasicrystalline) potential with two moiré patterns from top and bottom hBN interface each—may underlie some of these observations. This perspective is further supported by recent findings that such hysteretic responses depend sensitively on the twist angles in a similar hBN-graphene heterostructure with in-situ angle tunability [57].

Despite these findings, several open questions remain. In particular, it is still unclear where within the domain boundaries the localised electrons are found. The boundaries are structurally heterogeneous, containing segments with distinct lattice configurations such as domain walls and AAA sites [31]. While our data strongly suggest the presence of localised electrons within the boundaries, they do not reveal where exactly these states reside. Given that the moiré order is disrupted across the boundaries, localisation could plausibly occur anywhere along them—either throughout their entirety or only in certain segments. Future studies capable of mapping the local carrier density, for example using scanning capacitance or scanning single-electron transistor microscopy, could resolve this spatial distribution and provide further insight into its microscopic origin. It is also currently beyond the reach of our measurement whether this localisation is arising from pure structural random disorder or quasicrystalline order of moiré lattices (Supplementary Section V).

Beyond pursuing a fundamental understanding of this phenomenon, our findings also suggest practical applications. The robust and reproducible multistable hysteresis observed in our devices could be harnessed for multi-level non-volatile memory or neuromorphic computing components (**Extended Data Fig. 6**). Given the tunability afforded by twist angle, strain, and layer stacking in van der Waals heterostructures, our results underscore the potential of engineered aperiodic systems as functional quantum materials for next-generation electronic platforms.

# Methods

## Device fabrication

Helical trilayer graphene devices were fabricated by sequentially stacking graphene flakes encapsulated in hexagonal boron nitride (hBN). The process began with the mechanical exfoliation of pristine hBN (30–50 nm thick) and monolayer graphene onto a $SiO_2$/Si substrate (300 nm $SiO_2$). Candidate flakes were identified via optical contrast under an optical microscope. To minimise strain and wrinkles, we adopted a 'cut-and-stack' approach [58], in which graphene flakes were pre-cut using an atomic force microscope [59] rather than torn with hBN. The resulting pieces were assembled using a standard dry transfer method with a PDMS/polycarbonate (PC) stamp. The fully stacked heterostructures were then placed on a Si/$SiO_2$ substrate, ensuring that all stacking and drop-down steps were performed below 180 °C to avoid unintentional interlayer rotation. No thermal annealing was applied after stacking, in order to preserve the intended twist angles. A Ti/Au top gate was deposited on the heterostructure, and the Hall bar geometry was defined by electron-beam lithography followed by reactive ion etching. One-dimensional edge contacts were formed by depositing Cr/Au electrodes using electron-beam evaporation [60].

## Simulation of lattice configuration

Structural relaxation was performed using the Large-scale Atomic/Molecular Massively Parallel Simulator (LAMMPS) [61], an open-source package for real-space molecular modelling of systems with a large number of particles. Simulations were conducted on sufficiently large supercells to minimise edge effects and faithfully accommodate the unit cell geometry. The system was constructed by arranging three identical monolayer graphene sheets with a lattice constant of $a$ = 0.246 nm and a nearest-neighbour distance of $a_0$ = 0.142 nm. The top two layers were vertically shifted to form an ABA-stacked trilayer structure with an interlayer spacing of $c_0$ = 0.335 nm. Identical twist angles were then applied between layers 1 and 2 and between layers 2 and 3. Intralayer interactions were modelled using the Brenner potential [62], while interlayer interactions were described by the Kolmogorov–Crespi potential [63]. Following relaxation, AA-type stacking regions were identified between layers 1 and 2 ($AA_{12}$, marked by red dots) and between layer 2 and 3 ($AA_{23}$, blue dots). The resulting domain and domain boundary patterns are qualitatively consistent with previous studies [30,32].

## Measurements

Electrical transport measurements were performed in a dilution refrigerator with a base temperature of 20 mK. A standard four-probe measurement was employed, using lock-in detection with an AC excitation current of 1 nA at 17.777 Hz. Voltage signals were recorded using SR865A and SR830 lock-in amplifiers. The carrier density and displacement field were independently controlled by applying DC voltages to the Ti/Au top gate and the doped Si substrate, which served as a global bottom gate. Gate voltages were supplied using a GS200 voltage source. The charge carrier density, $n$, and displacement field, $D$, were determined using the following relations:

$$n = (c_{TG}V_{TG} + c_{BG}V_{BG})/e, \qquad D = (c_{BG}V_{BG} - c_{TG}V_{TG})/2\varepsilon_0$$

where $V_{TG}$ and $V_{BG}$ denote the voltages applied to the top and bottom gates, respectively; $\varepsilon_0$ is the vacuum permittivity; $c$ is the capacitance per unit area; and $e$ is the elementary charge.

**Twist angle determination**

The twist angle $\theta$ of each hTG device was determined from the Landau fan diagram by analysing the quantum Hall states that emerged in the system. The degeneracy of these states was identified through Hall measurements, enabling us to extract the superlattice carrier density $n_s$ from the slope of the corresponding fan features. Since the twist angle $\theta$ and $n_s$ are related via the following expression, $\theta$ can be obtained accordingly:

$$n_{\nu=\pm 4} = \pm 8\sin^2\theta/\sqrt{3}a^2$$

where the graphene lattice constant is $a = 0.246$ nm. The values of $\theta$ obtained through this procedure were further verified by examining their consistency with the Brown–Zak oscillation sequences.

**Electronic band calculation**

The band structure of hTG was simulated using the Bistritzer–MacDonald approach. For the helically twisted trilayer system, we consider three layers (labelled $l = 1,2,3$). In this structure, the outer layers are rotated in opposite directions by an angle $\theta$ while the middle layer reconstructs (shrinks), resulting in identical moiré lattices for the two bilayers. In each layer, the electron in the $p_z$ orbital at lattice site $i$ and sublattice $s$ is denoted by $|i, l, s\rangle$ (with spin neglected due to weak spin–orbit coupling), and its Bloch wave function in momentum space $\tilde{k}$ is written as $|\tilde{k}, l, s\rangle$. In the multilayer system, the full Hamiltonian couples different momentum states in different layers because the overall translational symmetry is broken. The sublattice position in each layer is modified by rotation and displacement as $\mathbf{r}_{ls} = R_{\theta_l} \cdot (\mathbf{r}_s + \mathbf{d}_l)$, with $\mathbf{d}_{1,2} = (0,0)$ and for layer 3, $\mathbf{d}_3$ chosen as either $a(\frac{\sqrt{3}}{2}, 0)$ or $a(0,1)$ for domain and domain wall configurations, respectively. Focusing on electrons near a specific valley $\mathbf{K}_l$, (writing $\tilde{\mathbf{k}} = \mathbf{K}_l + \mathbf{k}$, $\tilde{\mathbf{k}}' = \mathbf{K}_{l'} + \mathbf{k}'$), the intralayer Hamiltonian is expressed as

$$H_{l,l;\, s',s}(\mathbf{k}', \mathbf{k}) = -\delta_{\mathbf{k}',\mathbf{k}} v_f [e^{i\theta_l \sigma_z}(\sigma_x k_x + \sigma_y k_y)]_{s',s} + \delta_{s',s} V_l,$$

with Fermi velocity $v_f = 10^6$ m/s and $\sigma_{x,y}$ as Pauli matrices. The potential $V_l$ models an external electric field with $V_1 = -V$, $V_2 = 0$ and $V_3 = V$. The interlayer Hamiltonian is approximated by

$$H_{l',l;\, s',s}(\mathbf{k}', \mathbf{k}) = \sum_{i=1,2,3} (w + (|\mathbf{k} + \mathbf{K}_l - \mathbf{G}_l| - |\mathbf{K}_l|)\,\delta w)\, \delta_{l',l\pm 1}\, e^{i\,(\mathbf{G}^l_{l'} \cdot \mathbf{r}_{l's'} - \mathbf{G}^l_l \cdot \mathbf{r}_{ls})},$$

where to incorporate layer relaxation effect, the same-sublattice hopping is reduced by a factor $\kappa$, modifying the term by $[1 + (\kappa - 1)\delta_{s',s}]$. For our numerical calculations, we use the parameters $w = 110$ meV, $\delta w = -7$ meV nm, and $\kappa = 0.7$. A finite momentum mesh is generated connecting states through the interlayer coupling, and the full energy spectrum is obtained by diagonalizing the resultant Hamiltonian.

## Data availability

The data that support the findings of this study are available from the corresponding authors upon reasonable request.

## Acknowledgement

We thank Moon Jip Park for helpful discussions on the physical origin of the unusual transport. This work was supported by the National Research Foundation of Korea grants funded by the Ministry of Science and ICT (Grant Nos. RS-2025-23525425, RS-2020-NR049536, and RS-2023-00258359), the Institute for Basic Science of Korea (Grant No. IBS-R009- D1), SNU Core Center for Physical Property Measurements at Extreme Physical Conditions (Grant No. 2021R1A6C101B418), and Creative-Pioneering Researcher Program through Seoul National University. B.-J.Y. was supported by Samsung Science and Technology Foundation under project no. SSTF-BA2002-06, National Research Foundation of Korea (NRF) funded by the Korean government (MSIT), grant no. RS-2021-NR060087 and RS-2025-00562579, Global Research Development-Center (GRDC) Cooperative Hub Program through the NRF funded by the MSIT, grant no. RS-2023-00258359, Global-LAMP program of the NRF funded by the Ministry of Education, grant no. RS-2023-00301976. K. W. and T. T. acknowledge support from the JSPS KAKENHI (Grant Numbers 21H05233 and 23H02052) and World Premier International Research Center Initiative (WPI), MEXT, Japan.

## Author contribution

H.P., J.O. and J.J. conceived the project, H.P. and J.O. fabricated samples, performed measurements and analysed data, R.G., C.M. and B.-J.Y. performed DOS calculations and contributed to the interpretation of data, J.J. performed the lattice relaxation simulations, K.W. and T.T. provided hBN crystals, B.-J.Y. supervised the theoretical works and J.J. supervised the overall project. All authors provided inputs and contributed to the writing of the manuscript. H.P. and J.O. contributed equally to the work.

## Competing Interests

The authors declare no competing interests.

**Correspondence and requests for materials** should be addressed to Bohm-Jung Yang or Joonho Jang.

# Extended Data Figures

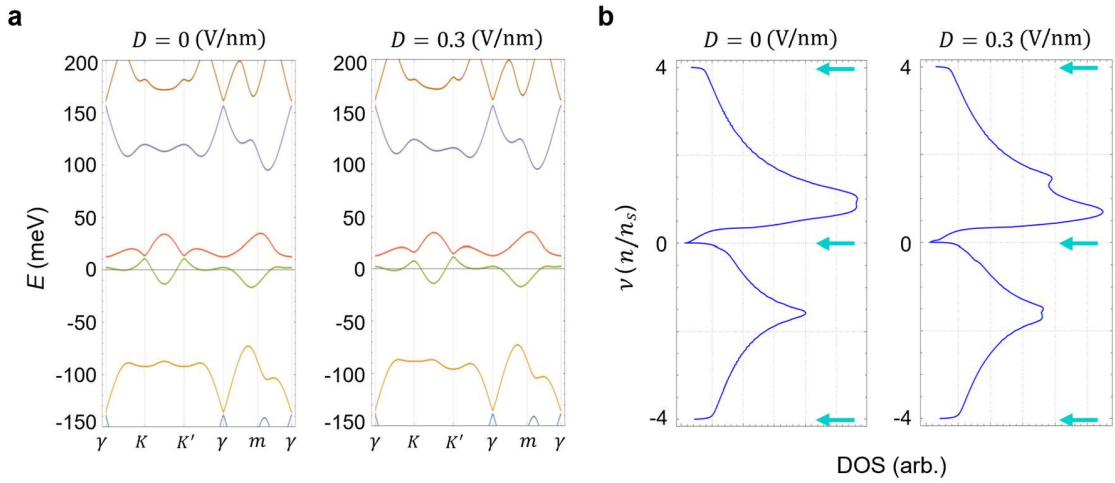

**Extended Data Fig. 1: Electronic band structure and density of states in relaxed domains**
**a,** Electronic band structures calculated for relaxed domains at a twist angle of 1.35°, for a single K valley, shown at displacement fields of 0 V nm$^{-1}$ (left) and 0.3 V nm$^{-1}$ (right). The left panel is reproduced from the first panel of Fig. 1b in the main text. **b,** Corresponding density of states (DOS) as a function of carrier density for displacement fields of 0 V nm$^{-1}$ (left) and 0.3 V nm$^{-1}$ (right). Cyan arrows mark the positions of energy gaps in each panel.

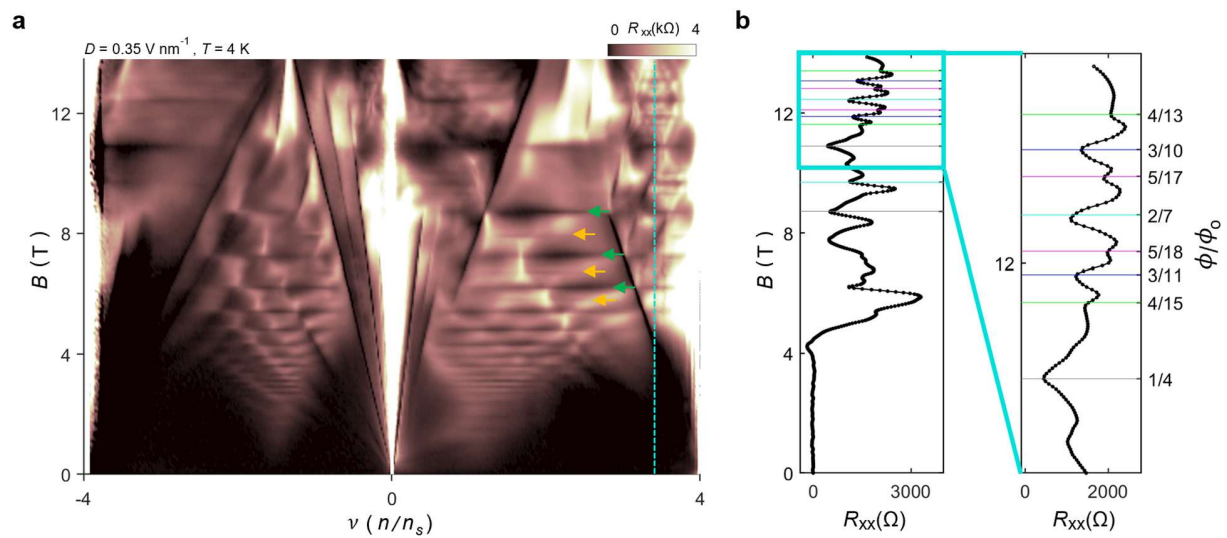

**Extended Data Fig. 2: High-order Brown–Zak (BZ) oscillations**
**a**, Landau fan diagram at $D = -0.35$ $V$ $nm^{-1}$ and 4 K, reproduced from Fig. 1e. Green and orange arrows indicate the first- and second-order BZ oscillations, which appear as horizontal low-resistance features. At higher magnetic fields, particularly near the region marked by the cyan dashed line, higher-order oscillations up to fifth order are observed. **b**, Line-cut of the resistance along the cyan dashed line in panel (a), plotted as a function of magnetic field. The full field range is shown on the left, with a magnified view of the region highlighted by the cyan box on the right. Each coloured horizontal line marks the position of a BZ oscillation, labelled by its corresponding order.

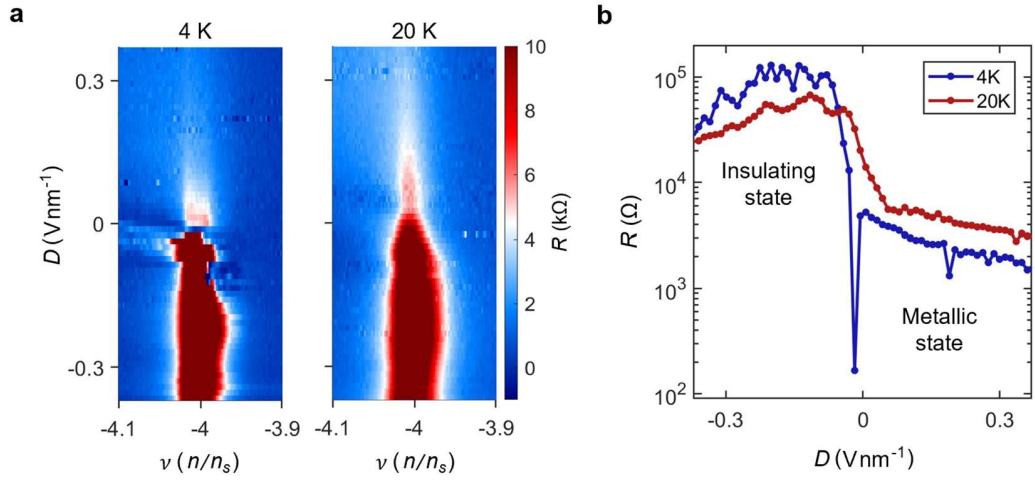

**Extended Data Fig. 3: Temperature-dependent resistance at $\nu = -4$**
**a**, Longitudinal resistance map near $\nu = -4$ as a function of $n$ and $D$ at 4 K (left) and 20 K (right). All data were acquired during upward gate voltage sweeps, as illustrated in the inset of Fig. 2b. **b**, Maximum resistance values for each $D$ from panel a, plotted on a logarithmic scale. Red and blue points correspond to measurements at 4 K and 20 K, respectively. The region left of the grey dashed line exhibits insulating-like behaviour, while the region to the right shows metallic-like behaviour.

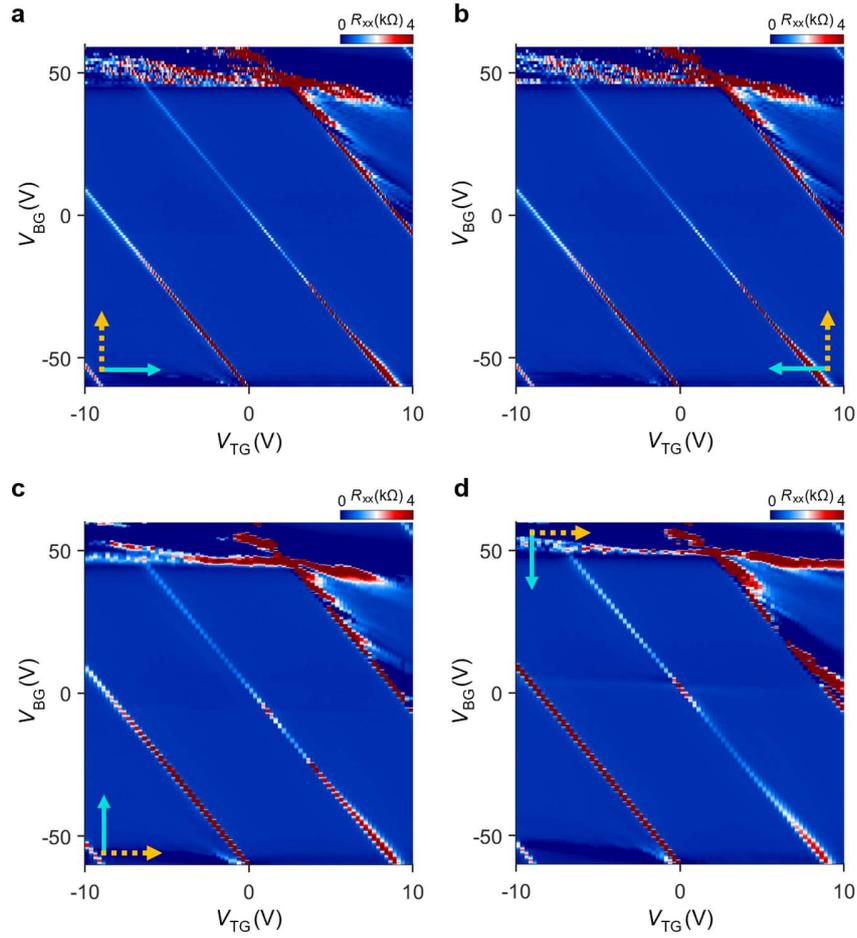

**Extended Data Fig. 4: Dual-gate sweep measurements with various sweep directions**
**a–b**, Longitudinal resistance measured at 4 K, plotted as a function of bottom gate voltage (slow axis, swept upward in both panels) and top gate voltage (fast axis), swept forward in panel a and backward in panel b. The two maps show minimal difference, indicating that the hysteresis is negligible when sweeping the top gate. **c–d**, Same measurement as in panels a–b, but with the top gate voltage as the slow axis (swept upward) and the bottom gate voltage as the fast axis, swept forward in panel c and backward in panel d. In contrast to panels a–b, a clear hysteretic difference emerges between panels c and d.

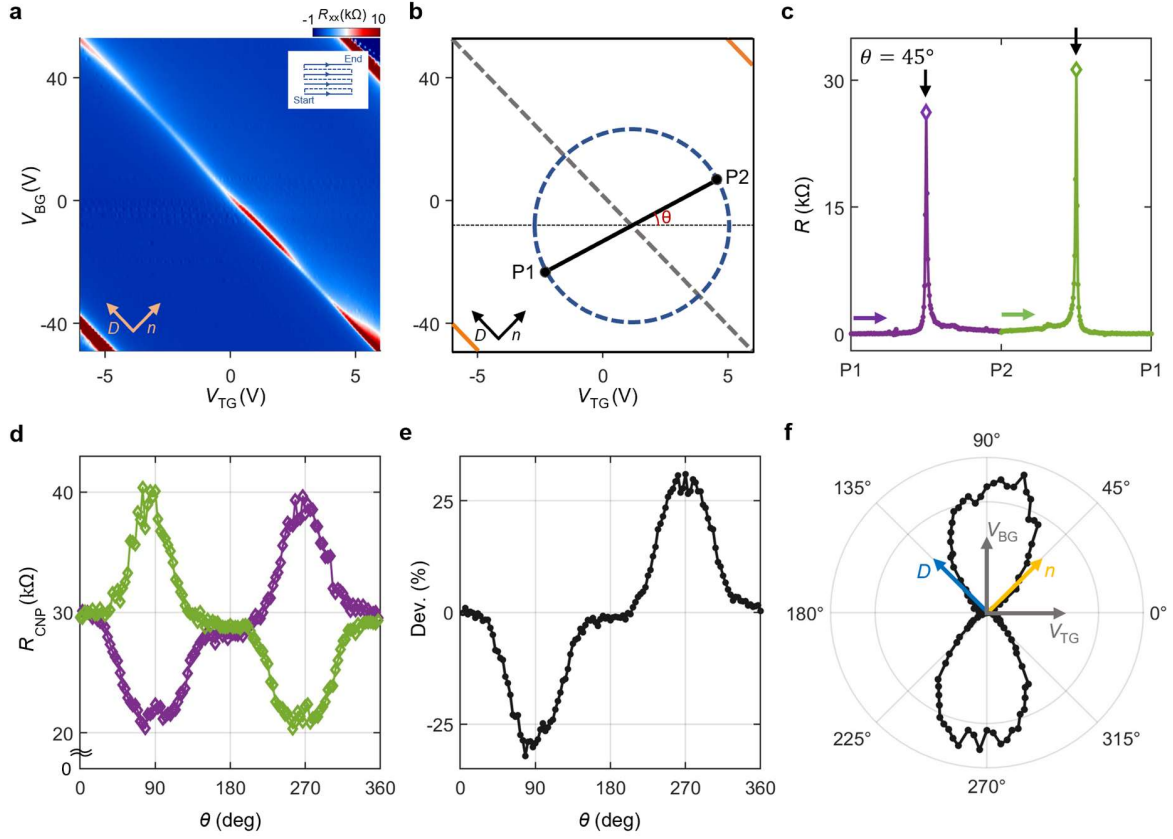

**Extended Data Fig. 5: Dependence of hysteresis on gate sweep angle**
**a**, Four-probe longitudinal resistance measurement as a function of $V_{TG}$ and $V_{BG}$ for the hTG2 device with a twist angle of 1.66°. Arrows in the inset indicate the gate sweep directions used during the measurement. **b**, Schematic highlighting the key features of panel a. The grey dashed line and orange solid lines mark the CNP and band insulators ($\nu = \pm 4$), respectively. The blue dotted ellipse defines a set of points satisfying: $((V_{TG}-1.25)/V_{TG,0})^2+((V_{BG}-9)/V_{BG,0})^2 = 1$, where $V_{BG,0} = 3.8$ V and $V_{BG,0} = 32.8$ V. Two points, P1 and P2, are selected along the ellipse such that the line connecting them passes through its centre and forms an angle $\theta$ with the horizontal axis. This angle defines the sweep trajectory used in panels c–f. **c**, Longitudinal resistance measured for a representative sweep at $\theta = 45°$, where gate voltages are swept from P1 to P2 (purple) and then from P2 to P1(green). **d**, Maximum resistance values extracted from forward and backward sweeps as $\theta$ is varied from 0° to 360°, shown purple and green, respectively. **e–f**, Relative resistance difference between the two sweeps in panel c, defined as $\Delta R/R_{avg}$, plotted as a function of $\theta$ in Cartesian coordinates (e) and in polar coordinates (f). A gate-asymmetric hysteresis is observed between the two sweep directions.

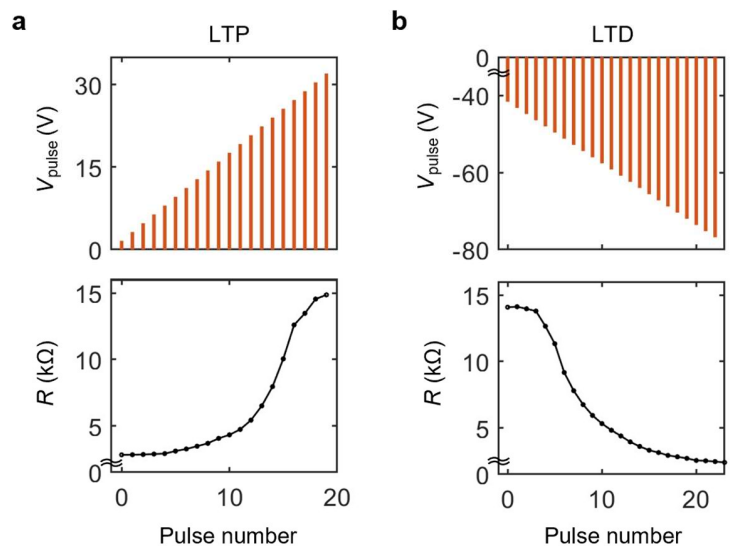

**Extended Data Fig. 6: Synaptic actions for neuromorphic computation**
**a–b,** Multifunctional synaptic responses of long-term potentiation (LTP) and long-term depression (LTD) as a function of pulse number, measured on the hTG2 device with a twist angle of 1.66°. The initial state is set at $(V_{TG}, V_{BG}) = (1.25\ V, -9\ V)$, a point on the CNP. From this state, additional pulse voltages are applied to $V_{BG}$. In both panels, the top subpanels show the applied pulse voltage ($V_{Pulse}$), and the bottom subpanels show the corresponding longitudinal resistance as a function of pulse number.

# Supplementary Information for:

# Evidence for electron localisation in a moiré-of-moiré superlattice

# List of Contents

# I. Additional data and supplementary analyses

Among the over 40 hTG devices we fabricated with twist angles ranging from 1.3° to 1.8°, only three exhibited a uniform twist angle over several micrometres. Each of these samples showed the hysteretic behaviour described in the main text, which was not observed in our alternating twisted trilayer device. The following are the measurement results for each sample.

## I.1. hTG1

Device hTG1, the main device used in the main text, has a twist angle of 1.35° and exhibits the following characteristics.
(1) Uniform twist angle over a target area and robustness against thermal cycling

To verify whether the device maintains a uniform twist angle over a large area, we first performed two-probe resistance measurements between adjacent electrodes. **Figure S1a** shows an OM image of the hTG1 device with electrode labels in red text. While sweeping the top gate, we measured the two-probe resistance between adjacent electrodes from electrode 1 to 8. The results are shown in **Figure S1b**. In each measurement, three resistance peaks are observed, and their positions are identical. Considering that the spacing between peaks corresponds to the number of electrons required to fill a moiré band, this result indicates that each two-probe measurement region is defined by a single twist angle. Thus, the region above the red dashed line in **Figure S1a** can be regarded as having a uniform twist angle.

Within this region, we performed four-probe measurements as functions of the dual gate using different configurations. In these two configurations, the current was applied from electrode 1 to 4, while the resistance was measured between electrodes 2 and 3 in one case (**Figure S1c**) and between electrodes 6 and 7 in the other (**Figure S1d**). In each measurement, the gate sweep was performed downward, as indicated in the insets of each figure. The two maps exhibit qualitatively similar features, further implying the uniformity of the sample. Notably, these measurements were taken during a different thermal cycle than the data presented in **Fig. 2a** of the main text, emphasizing that our results are also robust against thermal cycling. Both the data presented in the main text and the following figures were measured using the configuration shown in **Figure S1d**.

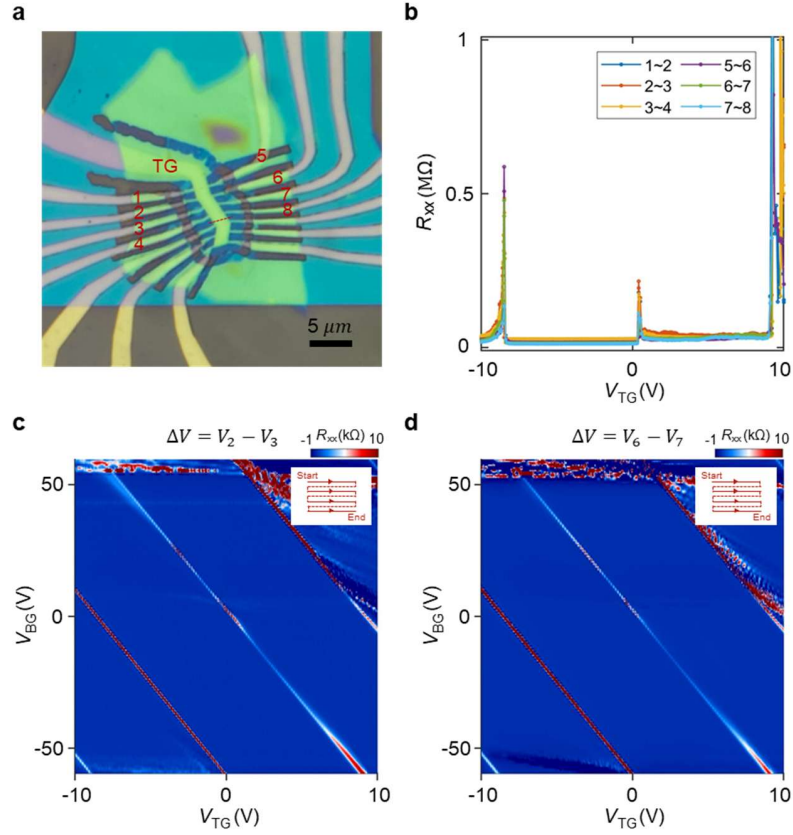

**Figure S1**
**a**, Optical image of the hTG1 device with a 5 μm scale bar. Electrode labels are shown in red text. **b**, Two-probe resistance measured at 50 mK under different configurations. **c–d**, Dual-gate sweep results obtained at 2.5 K using four-probe measurements between electrodes 2 and 3 (c), and between electrodes 3 and 4 (d). The current was applied from electrode 1 to 4. In both cases, the bottom gate was swept downward as the slow axis.

(2) Modification of the Hofstadter spectra by $D$ field

In **Fig. 1c** of the main text, we explained that the two local resistance minima observed in the resistance plot as a function of the $D$ field near the CNP correspond to topological band inversions predicted by band calculations. During these transitions, the Chern numbers of the two bands closest to the CNP change following the topological band inversion, which can modify the Hofstadter spectra. Considering the behaviour of the Hofstadter gap at the CNP under varying magnetic fields, it is well established that in systems with trivial topology, such gaps remain open and do not close [64]. Therefore, the gap closing observed here can be attributed to the intersecting Hofstadter subbands of distinct flavours, including spin and valley, and is consequently influenced by the topology of the involved bands [65].

For further investigation, we performed magnetoresistance measurements at $D = 0$ V nm$^{-1}$ and T = 1 K, with the results shown in **Figure S2a**. Notably, the gap at the CNP (high resistive region near $\nu = 0$) closes around 8 T, as can be seen in the cyan dotted box. This finding stands in stark contrast to the results obtained in the presence of a $D$ field, as shown in **Fig. 1d**, where the gap at the corresponding position remains open. To examine the dependence of the Hofstadter gap at the CNP on the $D$ field, we measured the resistance near the CNP while varying the $D$ field under a magnetic field of 11.5 T (**Figure S2b**). In this plot, the high resistance near the CNP sharply drops precisely at the positions of the local resistance minima in **Fig. 1c** (marked by pink arrows), which were predicted to correspond to topological transition points. This observation indirectly supports the presence of topological transitions at these positions.

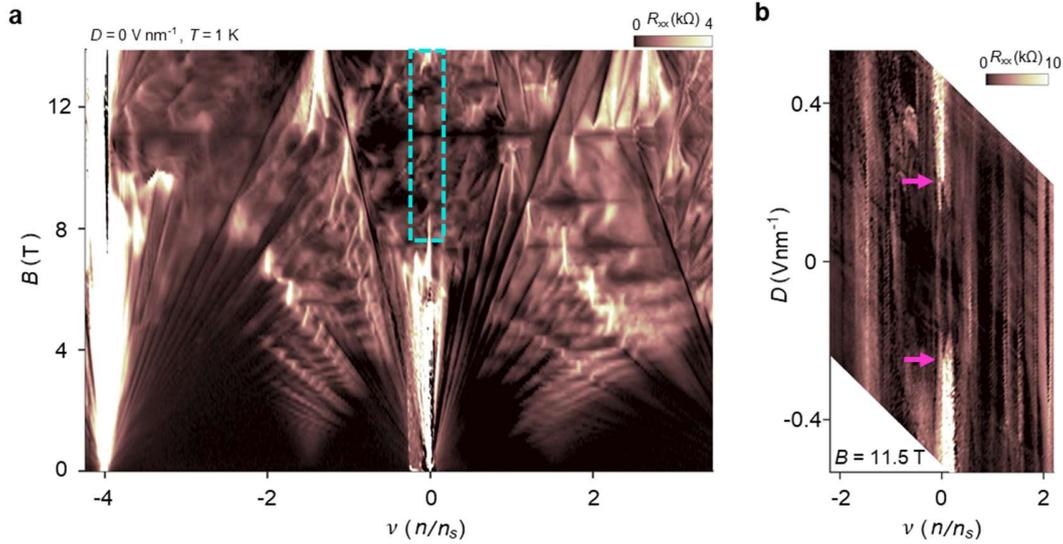

**Figure S2**
**a**, Landau fan diagrams of longitudinal resistance measured from the hTG1 at $D$ = 0 V nm$^{-1}$ and T = 1 K. The cyan dashed box highlights the region where the Hofstadter gap at the CNP closes. **b**, Resistance map near the CNP as a function of the $D$ field at 11.5 T. A sharp resistance drop is observed at the pink arrows, which mark the expected positions of topological band crossings.

## I.2. hTG2

hTG2 is the device used in **Extended Data Figs. 5** and **6** of the main text, with a twist angle of 1.66°. Due to its larger twist angle compared to hTG1, access to the band insulators is more limited. Still, it exhibits qualitatively the same hysteretic behaviour as hTG1, as detailed below.

(1) Basic characterisation and hysteretic response

The hTG2 device has the same geometry as hTG1. **Figure S3a** shows an OM image of the hTG2 device with electrode labels indicated in red. The four-probe measurement map in the *n-D* plane is shown in **Figure S3b**, obtained by applying current from electrode 1 to 4 and measuring the resistance between electrodes 2 and 3. During this measurement, the bottom gate was swept in the upward direction. Similar to the hTG1 device, the map reveals high-resistance vertical lines corresponding to $\nu$ = 0 and ±4. Additionally, the resistance at the CNP exhibits two local minima, as indicated by the pink arrows, which are predicted to correspond to topological band crossings. To investigate the hysteretic behaviour, we rescanned the resistance near the CNP in both the upward and downward sweep directions, as shown in **Figure S3c**. Consistent with the results in **Fig. 2** of the main text, hysteresis appears depending on the sweep direction, while the $D$ field positions of the dips near the CNP remain unchanged, as indicated by the pink dotted lines.

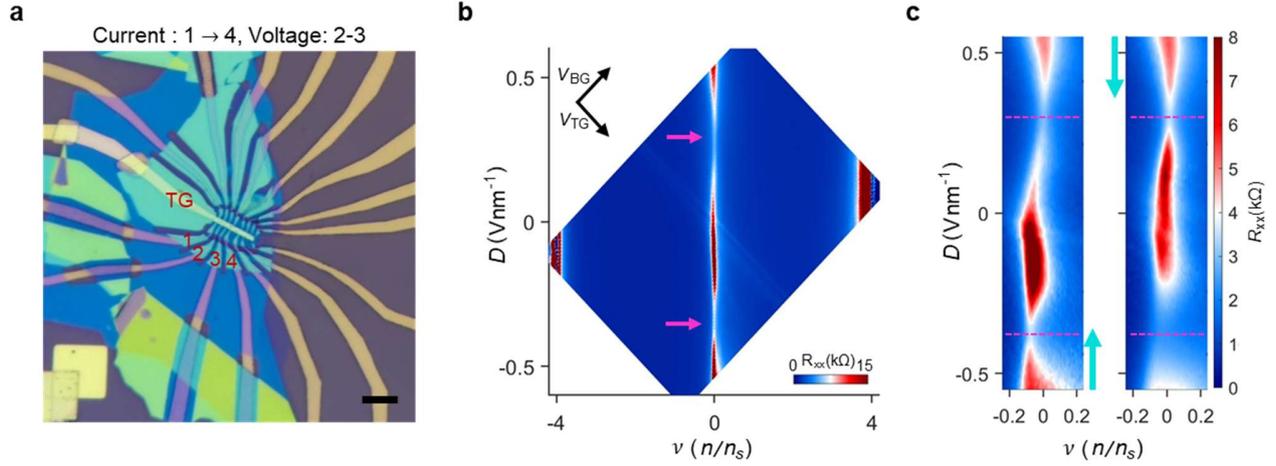

**Figure S3**
**a**, Optical image of the hTG2 device with a 10 μm scale bar. Electrode labels are indicated in red. **b**, Longitudinal resistance map of hTG2 as a function of $\nu$ and $D$ at 50 mK. Resistance was measured between electrodes 2 and 3 while applying current from electrode 1 to 4. As in the hTG1 device, the resistance near the CNP exhibits two local minima as the $D$ field is varied, marked by pink arrows. **c**, Hysteresis of resistance near the CNP under opposite sweep directions of the $D$ field. Measurements were performed at 4 K. A cyan arrow in each plot indicates the direction of the sweep.

(2) Magnetoresistance measurements at various $D$ field

This device also exhibits modifications to the Hofstadter spectra as a function of the $D$ field. Magnetoresistance measurements at 100 mK for $D = 0$ V nm$^{-1}$ and $D = -0.5$ V nm$^{-1}$ are shown in **Figure S4a** and **b**, respectively. Similar to the case of the hTG1 device, the Hofstadter gap at the CNP closes when $D = 0$ V nm$^{-1}$, as indicated by the cyan dashed box. In contrast, when $D$ is applied, the gap remains open up to 14 T, consistent with the behaviour observed in hTG1. Notably, the Landau fans on the electron side appear less developed than on the hole side, as seen in **Figure S4a**. This is likely due to the electron-side band becoming flatter as the twist angle of this sample approaches the magic angle of 1.8°.

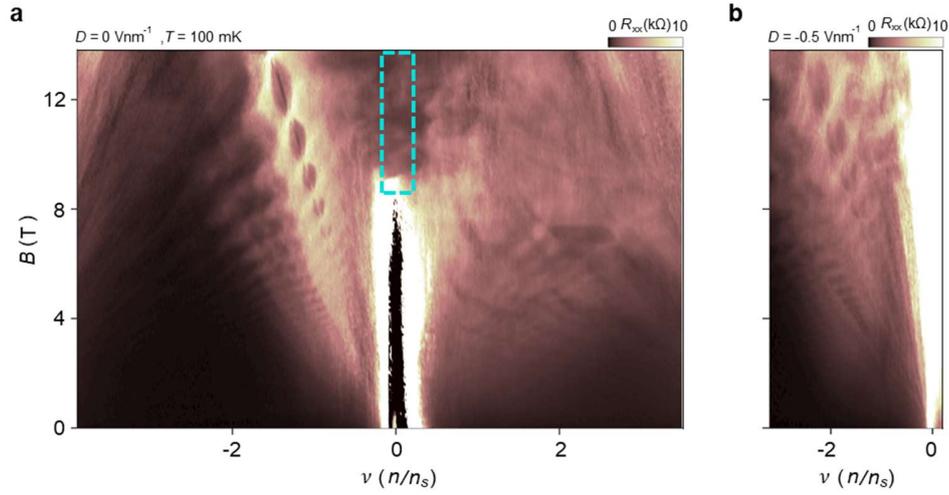

**Figure S4**
**a–b**, Landau fan diagrams of longitudinal resistance for hTG2 at $D = 0$ V nm$^{-1}$ (a) and $D = -0.5$ V nm$^{-1}$ (b) at T = 100 mK. When $D = 0$ V nm$^{-1}$, the Hofstadter gap is closed as indicated by the cyan dashed box, whereas at $D = -0.5$ V nm$^{-1}$, the corresponding region becomes insulating.

## I.3. hTG3

We have another sample that exhibits hysteretic behaviour, referred to as hTG3, with a twist angle of approximately 1.7° estimated from its geometric capacitance. The OM image of this device and the measurement configuration are shown in **Figure S5a**. The dual-gate sweep results in **Figure S5b** and **c** indicate that it qualitatively shares the same hysteretic behaviour as other hTG devices. We confirmed that this sample also exhibits hysteresis exclusively in the bottom gate sweep.

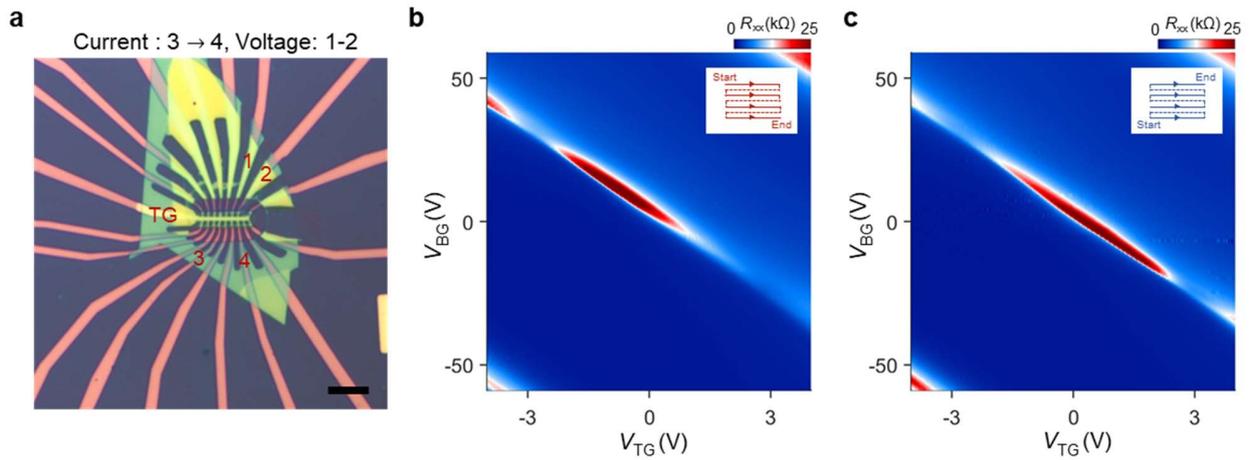

**Figure S5**
**a**, Optical image of the hTG3 device with a 10 μm scale bar. Electrode labels are indicated in red. **b–c**, Dual-gate resistance measurements at 1.8 K as functions of the bottom gate voltage (slow axis, swept upward in both panels) and the top gate voltage (fast axis), with $V_{TG}$ swept downward in (b) and upward in (c). Resistance was measured between electrodes 1 and 2 while applying current from electrode 3 to 4. As in other hTG devices, the resistance near the CNP exhibits distinctive hysteretic behaviour.

### I.4. aTG1

To confirm that the hysteretic behaviour observed in hTG devices originates from the system itself rather than being an artifact of the fabrication process, we fabricated an alternating twisted trilayer graphene device, referred to as aTG1, using the same fabrication process as the hTG devices. Unlike the hTG system, the alternating twisted trilayer graphene system has its two moiré patterns—one formed between the top and middle graphene layers and the other between the middle and bottom layers—perfectly aligned. As a result, commensurate moiré sites extend across the entire sample, preventing the formation of domain boundaries and the associated hysteresis.

**Figure S6a** shows an OM image of the aTG1 device, which shares the same geometry as the hTG devices. Among the many electrodes in the device, those in cleaner regions were selected to configure the measurement setup: current was applied from electrode 1 to 4 while measuring the resistance between electrodes 2 and 3. First, we grounded the top gate and performed magnetotransport measurements at 50 mK by sweeping the back gate from -60 V to 60 V, using the same range used for the hTG devices, with the resulting data shown in **Figure S6b**. The Landau fans are well developed at a few teslas, from which the twist angle was calculated to be 1.58°. We also performed measurements while varying the temperature from 50 mK to 2 K, as shown in **Figure S6c**, where a superconducting dome with a $T_c$ of approximately 700 mK was observed. These results indicate that the fabricated structure is a well-formed alternating twisted trilayer graphene system [66,67].

To examine whether hysteresis occurs, we set the temperature to 1 K, above the superconducting transition, and measured the resistance while sweeping the bottom gate forward and backward. The results, shown in **Figure S6d**, include green and purple traces, which correspond to the forward and backward sweeps, respectively. Importantly, no hysteresis is observed between the two sweeps, reinforcing that the distinct hysteretic behaviour observed in the hTG system originates from the system itself.

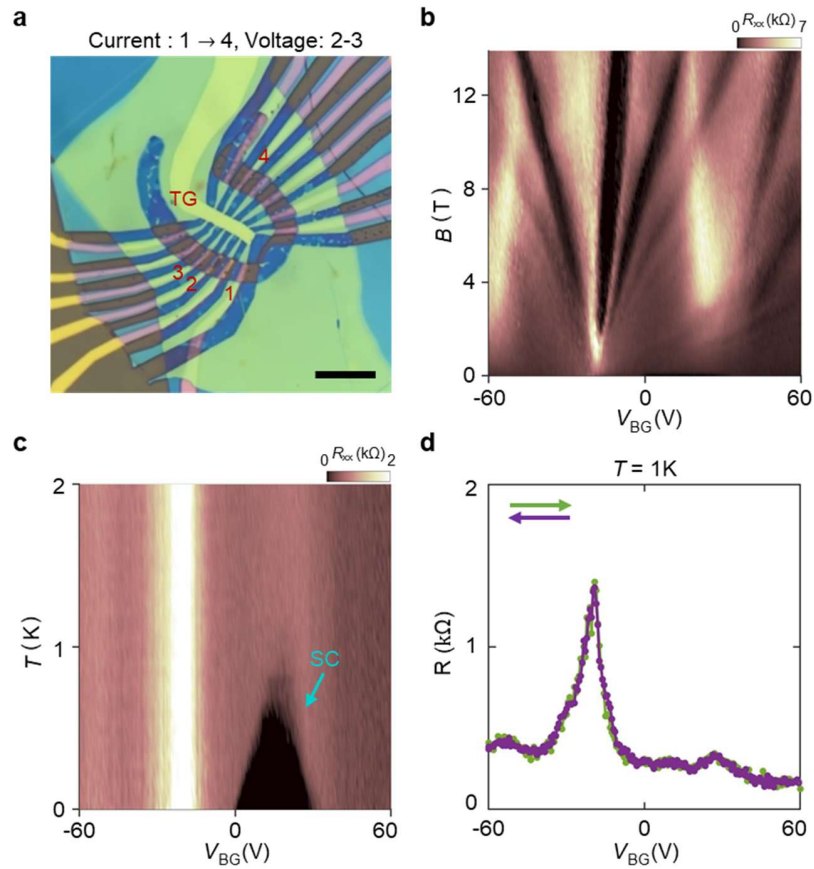

**Figure S6**
**a**, Optical image of aTG1 device with a 10 μm scale bar. The red annotations in the image indicate the electrode labels. **b**, Magnetoresistance measurement at 50 mK while sweeping the bottom gate from -60 V to 60 V, with the top gate held at 0 V. The twist angle extracted from the Landau fan is approximately 1.58°. **c**, Temperature-dependent resistance map at zero magnetic field. A superconducting dome with $T_c$ of approximately 700 mK exists in the bottom gate voltage range of 0 V to 30 V. **d**, Line cut plot at 1 K extracted from the map in (c). The green and purple data correspond to measurements taken while increasing and decreasing the gate voltage, respectively, indicating the absence of hysteresis.

## II. Electronic band structure calculations

In this section, we provide details regarding the electronic band structure calculations presented in the main text, along with supplementary data to aid further understanding.

### II.1. Details of band structure calculations using the continuum model

The band structure of hTG was simulated using the Bistritzer–MacDonald approach. A single graphene layer is described by the real-space lattice vectors $\mathbf{a}_1 = a\,(\sqrt{3}, 0)$ and $\mathbf{a}_2 = a\,(\frac{\sqrt{3}}{2}, \frac{3}{2})$, and the corresponding reciprocal lattice vectors $\mathbf{b}_1 = \frac{2\pi}{a}\left(\frac{1}{\sqrt{3}}, -\frac{1}{3}\right)$ and $\mathbf{b}_2 = \frac{2\pi}{a}\left(0, \frac{2}{3}\right)$ with $a = 1.42 \times 10^{-10}$ m. The two sublattices are located at $\mathbf{r}_1 = a(0,0)$ and $\mathbf{r}_2 = a\,(0,1)$. At half filling, Dirac cones appear at the inequivalent valleys $\mathbf{K}_0 = \frac{4\pi}{3\sqrt{3}a}$ and $\mathbf{K}_0' = -\frac{4\pi}{3\sqrt{3}a}$, where the low-energy excitations near $\mathbf{K}_0$ are described by the Dirac Hamiltonian

$$H_0 = v_\text{f}\,(\sigma_x k_x + \sigma_y k_y),$$

with Fermi velocity $v_f = 10^6$ m/s and $\sigma_{x,y}$ as Pauli matrices.

For the helically twisted trilayer system, we consider three layers (labelled $l = 1,2,3$). In this structure, the outer layers are rotated in opposite directions by an angle θ while the middle layer reconstructs (shrinks), resulting in identical moiré lattices for the two bilayers. In each layer, the electron in the $p_z$ orbital at lattice site $i$ and sublattice $s$ is denoted by $|i, l, s\rangle$ (with spin neglected due to weak spin–orbit coupling). Taking advantage of the translational symmetry within each layer, we introduce the momentum-space basis

$$|\tilde{\mathbf{k}}, l, s\rangle = (1/\sqrt{N}) \sum_i e^{i\,\tilde{\mathbf{k}} \cdot \mathbf{r}_l} |i, l, s\rangle,$$

where N is the number of lattice sites per layer. Assuming a tight-binding hopping $t(\mathbf{r}_J - \mathbf{r}_l)$ that depends on both distance and direction, the Hamiltonian in momentum space is

$$H_{l',l;s',s}(\tilde{\mathbf{k}}', \tilde{\mathbf{k}}) = 1/N \sum_{i,j} e^{i\,(\tilde{\mathbf{k}}' \cdot \mathbf{r}_J - \tilde{\mathbf{k}} \cdot \mathbf{r}_l)} t(\mathbf{r}_J - \mathbf{r}_l).$$

For an isolated layer, this Hamiltonian is block diagonal ($\tilde{\mathbf{k}}' = \tilde{\mathbf{k}}$) and yields Dirac cones centred at $\mathbf{K}_l$ and $\mathbf{K}_l'$. However, in the multilayer system, the full Hamiltonian couples states in different layers because the overall translational symmetry is broken. The sublattice position in each layer is modified by rotation and displacement as $\mathbf{r}_{ls} = R_{\theta_l} \cdot (\mathbf{r}_s + \mathbf{d}_l)$, with $\mathbf{d}_{1,2} = (0,0)$ and for layer 3, $\mathbf{d}_3$ chosen as either $a(\frac{\sqrt{3}}{2}, 0)$ or $a(0,1)$ for domain and domain wall configurations, respectively.

The interlayer hopping for the $p_z$ orbitals is expressed in Fourier space as

$$t(\mathbf{r}) = \int d^2q\, t(|\mathbf{q}|)\, e^{-i\mathbf{q} \cdot \mathbf{r}},$$

and by writing the position as $\mathbf{r}_l = \mathbf{R}_l + \mathbf{r}_{ls}$, the momentum-space Hamiltonian becomes

$$H_{l',l;\,s',s}(\tilde{\mathbf{k}}',\tilde{\mathbf{k}}) = 1/N \int d^2q\, t(|\mathbf{q}|) \sum_j \sum_i e^{i(\tilde{\mathbf{k}}'\cdot\mathbf{R}_J+\tilde{\mathbf{k}}'\cdot\mathbf{r}_{l's'}-\tilde{\mathbf{k}}\cdot\mathbf{R}_I-\tilde{\mathbf{k}}\cdot\mathbf{r}_{ls})} e^{-i\mathbf{q}\cdot(\mathbf{R}_J+\mathbf{r}_{l's'}-\mathbf{R}_I-\mathbf{r}_{ls})}.$$

After summing over lattice sites, delta functions enforce momentum conservation modulo reciprocal lattice vectors. Focusing on electrons near a specific valley (writing $\tilde{\mathbf{k}} = \mathbf{K}_l + \mathbf{k}$, $\tilde{\mathbf{k}}' = \mathbf{K}_{l'}' + \mathbf{k}'$), the interlayer coupling is written as

$$H_{l',l;\,s',s}(\tilde{\mathbf{k}}',\tilde{\mathbf{k}}) = 1/\Omega \sum_{\mathbf{G}_l,\mathbf{G}_{l'}} t(|\mathbf{k}+\mathbf{K}_l-\mathbf{G}_l|)\, \delta_{\mathbf{k}'-\mathbf{k},\mathbf{K}_l-\mathbf{K}_{l'}+\mathbf{G}_{l'}-\mathbf{G}_l}\, e^{i(\mathbf{G}_{l'}\cdot\mathbf{r}_{l's'}-\mathbf{G}_l\cdot\mathbf{r}_{ls})}.$$

Since $t(|q|)$ decays rapidly with increasing $q$, only three sets of reciprocal lattice vectors (indexed by $i = 1,2,3$) contribute significantly. We approximate

$$t(|\mathbf{k}+\mathbf{K}_l-\mathbf{G}_l|)/\Omega = w + (|\mathbf{k}+\mathbf{K}_l-\mathbf{G}_l|-|\mathbf{K}_l|)\,\delta w,$$

with $w = t(|\mathbf{K}_l|)$ and $\delta w = \frac{dt(q)}{dq}|_{q=|\mathbf{K}_l|}$. Then, the interlayer Hamiltonian is then given by

$$H_{l',l;\,s',s}(\mathbf{k}',\mathbf{k}) = \sum_{i=1,2,3}(w + (|\mathbf{k}+\mathbf{K}_l-\mathbf{G}_l|-|\mathbf{K}_l|)\,\delta w)\,\delta_{l',l\pm1}\, e^{i(\mathbf{G}_{l'}^l\cdot\mathbf{r}_{l's'}-\mathbf{G}_l^l\cdot\mathbf{r}_{ls})}.$$

To incorporate layer relaxation effect, the same-sublattice hopping is reduced by a factor $\kappa$, modifying the term by $[1+(\kappa-1)\delta_{s',s}]$ (with $\kappa = 0$ defining the chiral limit).

The intralayer Hamiltonian is expressed as

$$H_{l,l;\,s',s}(\mathbf{k}',\mathbf{k}) = -\delta_{\mathbf{k}',\mathbf{k}}\, v_f\, [e^{i\theta_l\sigma_z}(\sigma_x k_x + \sigma_y k_y)]_{s',s} + \delta_{s',s} V_l,$$

where the potential $V_l$ models an external electric field with $V_1 = -V$, $V_2 = 0$, and $V_3 = V$.

For our numerical calculations, we use the parameters $w = 110$ meV, $\delta w = -7$ (in suitable units), and $\kappa = 0.7$. A momentum mesh is generated connecting states through the interlayer coupling, and the full energy spectrum is obtained by diagonalizing the resultant Hamiltonian. This approach captures the low-energy physics of the twisted trilayer graphene system while retaining the essential details of the original tight-binding and Fourier analysis.

## II.2. Band structures of domain and domain wall regions

When considering the relaxed lattice structure, the system consists of domains surrounded by aperiodic domain boundaries, as illustrated in the panels of **Figure S7a**. As shown in previous references [32,33,35] and our simulations [30], the system contains two types of domains related by $C_{2z}$ symmetry, commonly referred to as h-domain and $\bar{h}$-domain (see right panel). Each domain is characterized by its own displacement vector, $d_1$ and $d_2$, which are respectively defined by the relative shift between moiré sites associated with different moiré patterns, as illustrated in the first and third panels of **Figure S7b**. Since these domains exhibit well-defined periodicity, as indicated by the wavelength $\lambda_m$ in the first panel, their band structure can be computed within the continuum approximation.

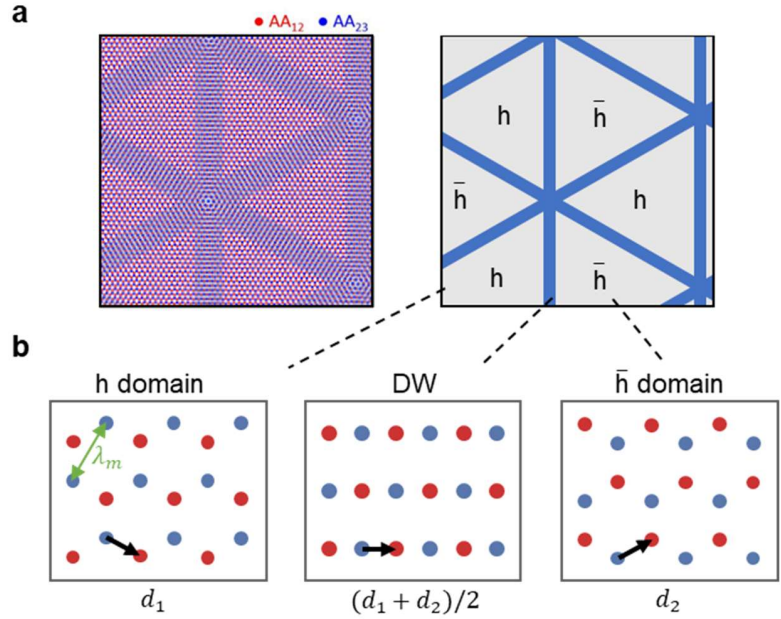

**Figure S7**
**a**, Simulated lattice configuration (left) and corresponding schematic representation (right). In the left panel, the red and blue dots represent the AA sites of two distinct moiré patterns. The right panel shows h-domain and $\bar{h}$-domain regions; the blue lines mark the domain boundaries. **b**, Schematics of the local moiré configuration for each region. The domain regions are characterized by displacement vectors d, which represent the relative shift between the red and blue lattice point groups. The displacement vector of the domain wall (DW) region is defined as the average of those of the two adjacent domains.

In contrast to the periodic domains, the domain boundaries possess an aperiodic distribution of moiré sites due to moiré relaxation. For the band structure calculations, we focused on a representative part of the boundary region, referred to here as the domain wall (DW), which corresponds to the transitional area between two adjacent domains. This DW region was approximated using the average of the two adjacent displacement vectors, $(\mathbf{d_1}+\mathbf{d_2})/2$, as illustrated in the middle panel of **Figure S7b**.

The electronic structures of the h-domain and DWs, calculated for a single valley at a twist angle of 1.35° under this assumption, are shown in **Figure S8a**, and the corresponding density of states (DOS) as a function of carrier density is presented in **Figure S8b**. As indicated by the cyan arrows, both regions exhibit a suppression of DOS at $\nu = 0$ and $\pm 4$. This result supports our interpretation in **Fig. 3e** that the domain boundaries become resistive at similar filling factors, consistent with the suppression of DOS. Notably, our calculations suggest the presence of valley-protected states in the DW region at $\nu = \pm 4$. If these dissipationless channels remain intact at the device scale, the measured resistance should be lower than the von Klitzing constant, $h/e^2$, which corresponds to a perfectly dissipationless channel. However, as shown in our data, the measured resistance at these fillings is significantly higher than this value, with some regions exceeding 100 kΩ, likely due to valley mixing induced by lattice disorder in the DW, which can open a gap.

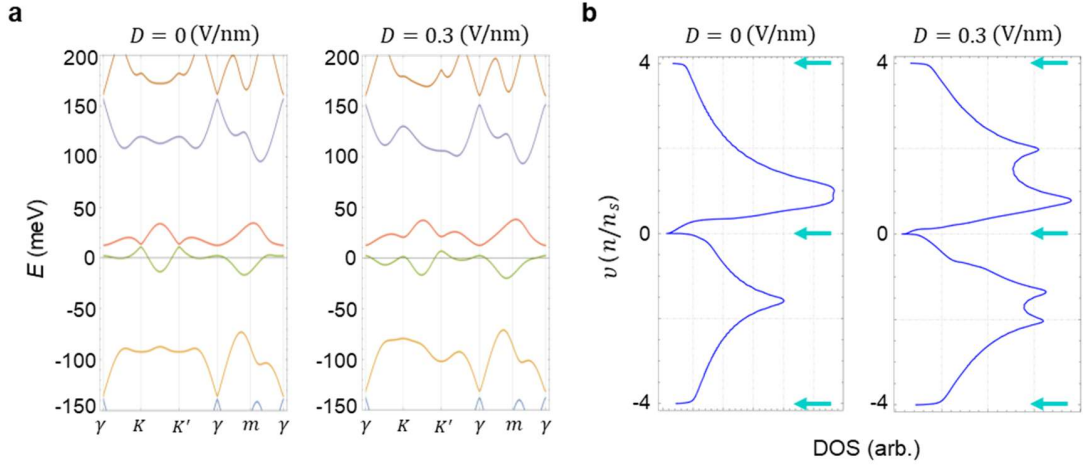

**Figure S8**
**a**, Band structures calculated for the h-domain (left) and the DW region (right) for the K valley in a system with a twist angle of 1.35°.
**b**, DOS plots as a function of carrier density, derived from the band structure calculations, for the h-domain (left) and the DW region (right). The cyan arrows in each panel indicate the positions of $\nu$ = -4, 0, and 4, where the DOS is locally suppressed.

## II.3. Band structure calculations at varying displacement fields

To examine the changes in the band structure of the domain regions under a $D$ field, we varied the $D$ values in our calculations and computed the corresponding electronic band structures for a single K valley, some of which are shown in **Figure S9a**. When $D = 0$ V nm$^{-1}$ (as shown in the first panel), the bands around the CNP are isolated and exhibit Chern numbers of +2 and -1, respectively. Notably, even when $D = 0$, each domain does not preserve $C_{2z}$ symmetry and, as a result, can host nontrivial topology, which gives rise to these Chern numbers. As the $D$ value increases, the gap between the two isolated bands around the CNP gradually decreases until they touch, as indicated by the pink circle in the second panel, and then increases again. Remarkably, as shown in the third panel, once the bands cross and separate their Chern numbers change from +2 and -1 to +1 and 0, confirming that this touching process represents a topological band inversion. The extracted band gap at the CNP during this process is plotted in **Figure S9b**. The regions marked by pink arrows in the plot indicate where the two bands cross, which we associate with the local resistance minima observed in **Fig. 1c**.

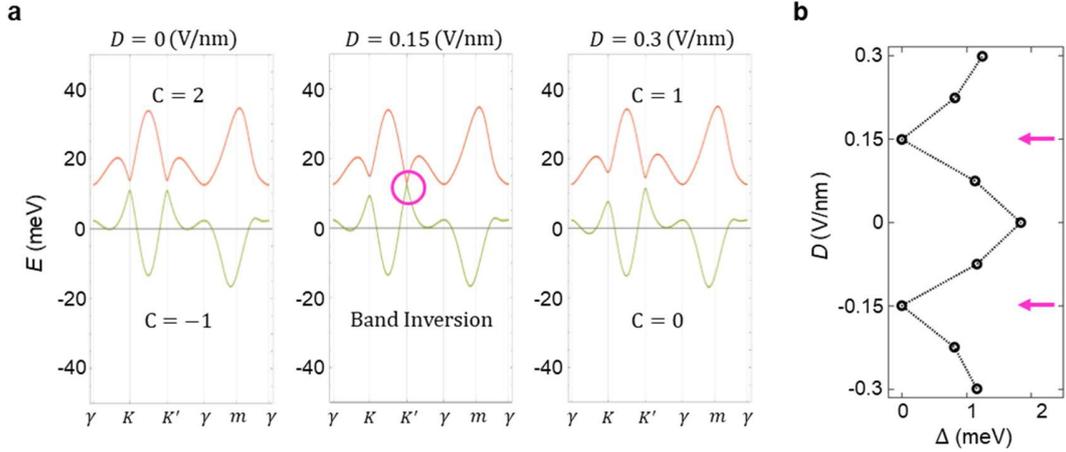

**Figure S9**
a, Calculated band structures for the domain region at a twist angle of 1.35°, shown at varying displacement field values. The left, middle, and right panels correspond to $D = 0$, $0.15$, and $0.3$ V nm$^{-1}$, respectively. As $D$ increases, the two bands near the CNP undergo a topological band inversion (highlighted by pink circle), resulting in a change in the Chern number of each band. b, Band gap at the CNP as a function of $D$. The regions marked by pink arrows indicate the locations of topological band inversions.

**Figure S10** presents the calculated band structures over an extended energy range, in contrast to **Figure S9**, which focuses on bands near the CNP. By expanding the vertical axis, additional higher-energy bands become visible, offering a broader view of the spectrum. Notably, sizeable band gaps—on the order of several tens of meV—remain open at filling factors $\nu = \pm 4$, even as the displacement field $D$ is varied over a broad range from $-0.6$ to $0.6$ V nm$^{-1}$. This persistence highlights the stability of the insulating states at these fillings against $D$ field modulation.

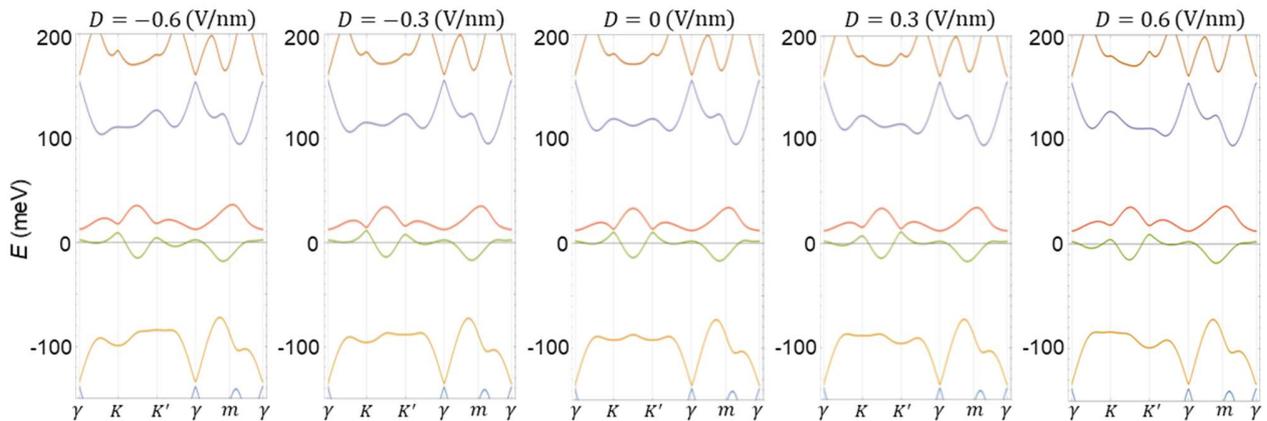

**Figure S10**
Band structure calculations for a single domain region at a twist angle of 1.35°, shown for five different displacement field values: $-0.6$, $-0.3$, $0$, $0.3$ and $0.6$ V nm$^{-1}$ (from left to right). Each panel displays the electronic structure over an extended energy range.

# III. Supplementary data on hysteretic transport behaviour

This section provides supplementary data and interpretations related to the hysteretic transport behaviour of hTG, complementing the analysis presented in the main text.

## III.1. Parallel resistance model interpretation of hysteresis

We model the total resistance of our hTG system in terms of its two constituent components—those from the domains and the domain boundaries—which are modelled as connected in parallel, as depicted schematically in **Fig. 2f**. As shown in **Figure S11a**, the first panel (grey dashed line) schematically represents the resistance of the domains, $R_{\text{Domain}}$, and the second panel (blue dashed line) shows that of the domain boundaries, $R_{\text{Boundaries}}$. These schematics are reconstructed based on the data presented in **Fig. 2**. Because $R_{\text{Boundaries}}$ is regarded as exhibiting hysteretic behaviour, its resistance peak can shift depending on the direction of the gate voltage sweep—this possibility is indicated by an arrow in the second panel. Under the assumption of a parallel configuration, the total resistance is given by

$$1/R_{\text{total}} = 1/R_{\text{Domain}} + 1/R_{\text{Boundaries}}.$$

The two rightmost panels of **Figure S11a** show the resulting $R_{\text{total}}$ traces (red solid lines) for two distinct cases of $R_{\text{Boundaries}}$, calculated using this relation. As seen in the third panel, the resulting resistance trace closely resembles the line-cuts in **Fig. 2g**, suggesting that the data can be naturally understood within this parallel configuration model. Conversely, with a measured $R_{\text{total}}$, one can use this model to estimate the two constituent resistances. In the main text, we applied this approach by assuming a maximum peak value of 1000 k$\Omega$ for $R_{\text{Boundaries}}$, used as a free parameter required for separating the resistance components.

Additional support for this model is provided by the situation in which the resistive packet aligns with the dips in the non-hysteretic domain resistance. Under these circumstances, the model predicts that the packet should split around the band inversion point, as highlighted by the purple arrows in the rightmost panel of **Figure S11a**. This prediction is verified by experiment: when the gate voltage is tuned such that the resistive packet coincides with one of the band inversions, it indeed separates into two smaller packets, as shown in the resistance map in **Figure S11b**. The corresponding line-cut, presented in **Figure S11c**, also exhibits a shape closely resembling the schematic in the rightmost panel of **Figure S11a**, thereby reinforcing the validity of our model.

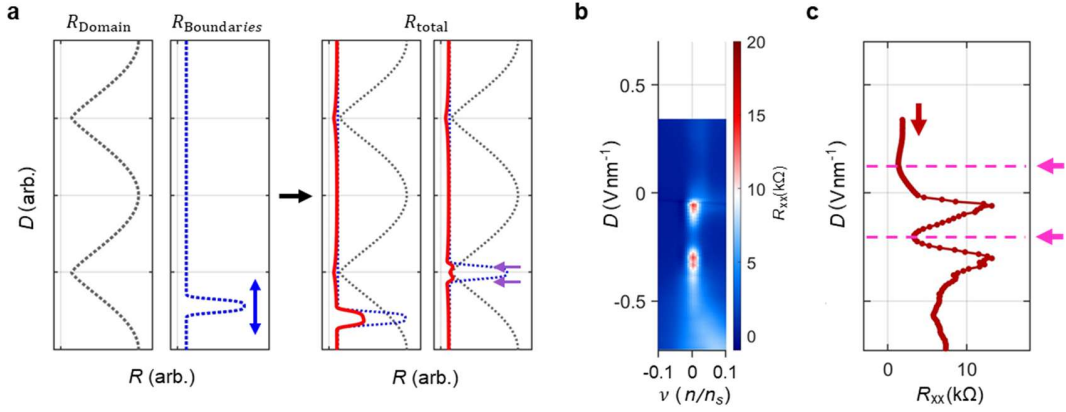

**Figure S11**
**a**, Schematic illustrations of the resistance responses to varying $D$ fields within different regions of the hTG system. The leftmost panel shows the resistance of the domain region ($R_{Domain}$, grey dotted line), and the second panel shows that of the domain boundaries ($R_{Boundaries}$, blue dotted line) with an arrow indicating the potential shift in resistance depending on the sweep direction. The two rightmost panels display the total resistance ($R_{total}$, red solid line), calculated by combining the two components in parallel according to the relation $1/R_{total} = 1/R_{Domain} + 1/R_{Boundaries}$, for two different cases, with the individual components also shown in the same panels for comparison. **b**, Longitudinal resistance map when the resistive packet is positioned at the band inversion point. **c**, Maximum resistance value at each $D$ field, extracted from b. The observed splitting of the resistive packet at the position marked by the pink arrow is consistent with the model prediction shown in a.

## III.2. Evolution of the resistive packet position in the *n-D* plane

In the main text, we highlighted that the Hall resistance ($R_{xy}$) not only exhibits hysteresis but also deviates from the $\nu = 0$ line. In contrast, the resistive packet is found to evolve along the $\nu = 0$ trajectory, as demonstrated in the data presented here. **Figure S12a** shows $R_{xx}$ maps acquired simultaneously with the $R_{xy}$ maps presented in **Fig. 3** of the main text. The left and right panels correspond to upward and downward gate sweeps, respectively, with blue arrows indicating the sweep direction. All maps were measured at 4 K and ±0.2 T, and a pink dotted line is overlaid to indicate the $\nu = 0$ trajectory. In these maps, the high-resistance features near the CNP do not follow this line and instead appear slightly shifted with a finite slope. The resistive packet itself, however, consistently follows the $\nu = 0$ line. To further examine this behaviour, we additionally acquired $R_{xx}$ maps while varying the gate sweep range, as shown in **Figure S12b**. These measurements were taken at 4 K and ±0.1 T. Blue arrows in each panel indicate both the direction of the sweep and the initial $D$ field value. Each map again includes a pink dotted line representing the $\nu = 0$ position. Across all conditions, the high-resistance features near the CNP remain slightly deviated from the pink dotted line, yet the resistive packet continues to appear along it. These results confirm that the resistive packet consistently emerges along the $\nu = 0$ line.

This behaviour is consistent with our interpretation of the hysteresis presented in **Fig. 3** of the main text. The non-hysteretic domains have their CNP aligned with $\nu = 0$, and the resistive packet arises when the CNP of the hysteretic domain boundaries intersects this trace. Consequently, the packet should appear along the $\nu = 0$ line. The apparent deviation of the high-resistance features near the CNP can also be understood in this context: when the CNP of the domain boundaries deviates from $\nu = 0$, the high-resistance feature in the total resistance appears between the CNP positions of the domains and the domain boundaries, and thus deviates from the $\nu = 0$ line as well.

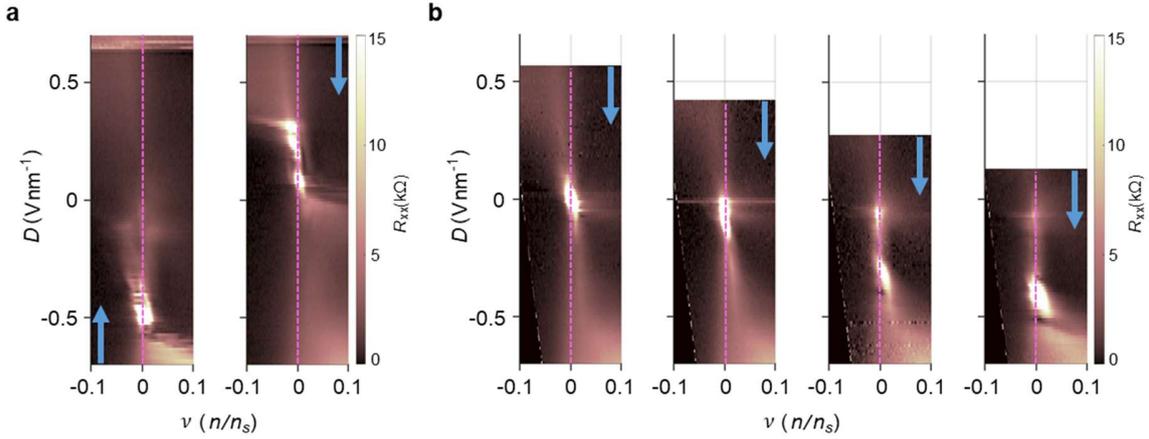

**Figure S12**
**a-b**, Longitudinal resistance maps plotted in the $n$–$D$ plane near the CNP, measured at 4 K. Data in panels a and b were acquired under magnetic fields of ±0.2 T and ±0.1 T, respectively. Blue arrows indicate the gate sweep direction and the initial $D$ field value for each sweep. Pink dashed lines mark the $\nu = 0$ position in all maps.

### III.3. Hall resistance behaviour at ±4 filling

While the main text focused on the hysteretic Hall response near the CNP (**Fig. 3a**), similar behaviour is also observed near $\nu = \pm 4$. **Figure S13a,b** shows $R_{xy}$ maps near $\nu = -4$ and $+4$, respectively, measured at 4 K. Each displays upward (left) and downward (right) gate sweeps, with pronounced sweep-direction-dependent responses analogous to those at the CNP. However, due to the highly resistive features at exactly $\nu = \pm 4$, the Hall signals are too noisy to reliably extract the carrier density of the domain boundaries using the method employed in the main text. We therefore analysed line cuts slightly offset from these fillings—specifically at $\nu = -3.973$ and $+3.973$. The longitudinal resistance along these lines reveals regions with high resistance (exceeding 100 kΩ), indicating that the domain remains highly resistive along these trajectories. Therefore, consistent with the analysis presented in Supplementary Section IV, the Hall resistance along these lines can still be used to estimate the carrier density in the domain boundaries. The extracted carrier densities at each filling are summarised in **Fig. 3e**.

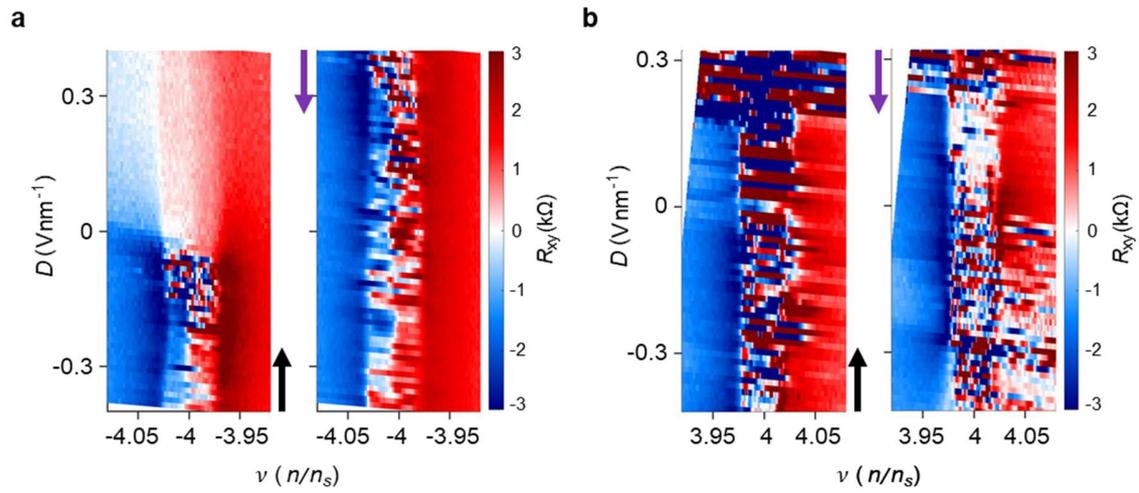

**Figure S13**
**a-b**, Hall resistance maps plotted in the $n$–$D$ plane near $\nu = -4$ (a) and $\nu = +4$ (b), each showing measurements taken during upward (left) and downward (right) gate sweeps. All data were acquired at 4 K. These datasets reveal pronounced hysteretic behaviour, analogous to that observed near the CNP.

## IV. Finite-element analysis of Hall response

To investigate the Hall response in the hTG system containing both domains and domain boundaries, we carried out finite-element simulations using the COMSOL Multiphysics software. The calculations were performed using the 2D Electrostatics interface in stationary mode. We modelled a representative geometry consisting of a periodic array of domains and domain walls, confined to a rectangular region of 12 μm in length and 3 μm in width, as shown in **Figure S14a**. Within this area, the supermoiré periodicity is 450 nm and the domain wall width is 65 nm. The chosen moiré length of 450 nm roughly corresponds to that of hTG with a twist angle of ~1.35°.

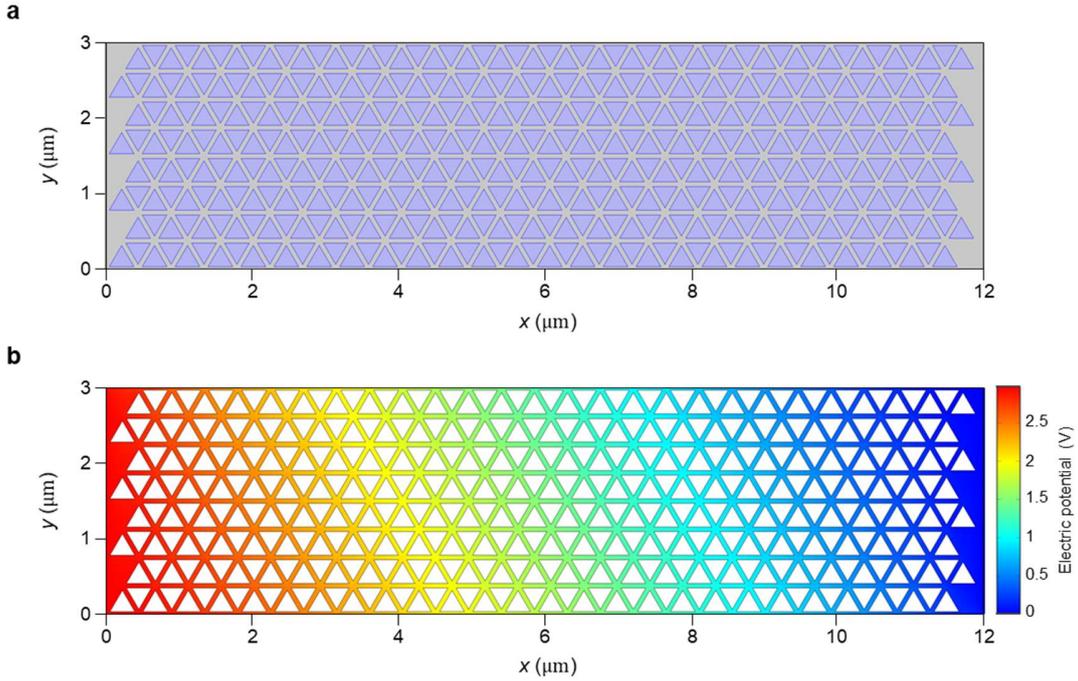

**Figure S14**
**a**, Geometry used for COMSOL simulation, consisting of a periodic array of triangular domains (purple) embedded in a background representing the domain boundaries (grey). **b**, Simulated spatial distribution of the electric potential, visualised only within the domain boundaries, under a 1 A current applied from left to right. The conductivity tensor in the domain boundaries is set to σ = [1 2; –2 1], corresponding to μB = 2, while the domains have a conductivity $10^{-5}$ times lower.

For the boundary conditions, a constant current of 1 A was applied along the left edge ($x = 0$) while the right edge ($x = 12$ μm) was grounded, maintaining an equipotential surface on the current injection side. To emulate the effect of an out-of-plane magnetic field, we introduced an off-diagonal term in the conductivity tensor. In the domain boundaries, the conductivity tensor was set to σ = [1 2; -2 1], based on an assumed mobility of 100,000 cm²/Vs under a magnetic field of 0.2 T, such that μB = 2. In the initial setup, the domains were treated as highly insulating regions, with their conductivity set to $10^{-5}$ times that of the boundaries. For simplicity, we set the reference conductivity $\sigma_0 = 1$. These parameter values were set for illustrative purposes, as the resulting potential profile depends only on their relative scales. **Figure S14b** shows the resulting spatial distribution of the electric potential, visualised only within the domain boundaries. The calculated potential, despite being limited to the domain boundaries, appears qualitatively similar to that of a homogeneous system.

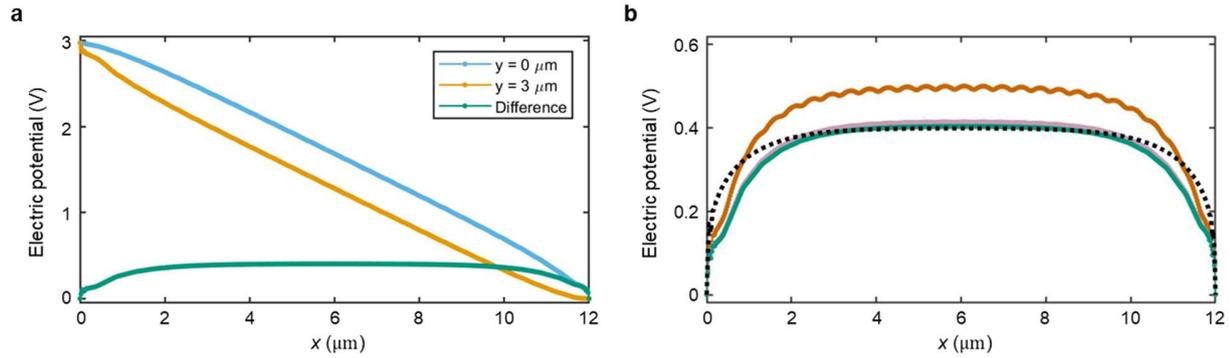

**Figure S15**
**a**, Electric potential along the bottom ($y = 0$ μm, light blue) and top ($y = 3$ μm, orange) edges of the simulated geometry, and their difference (green), corresponding to the Hall voltage. The potential difference remains nearly constant in the central region, away from the sample edges. **b**, Calculated Hall voltages for different conductivity ratios between the domains and the domain boundaries: 0.1 (brown), 0.01 (lavender grey), and $10^{-5}$ (green, same as in panel a). The black dashed line corresponds to the result for a homogeneous system without domain structure.

Since the Hall voltage is defined as the potential difference across the edges perpendicular to the current flow, we examined it by extracting the electric potential along the top ($y = 3$ μm) and bottom ($y = 0$ μm) edges of the simulated region in **Figure S14b**. As shown in **Figure S15a**, the light blue and orange traces represent the potential along $y = 0$ and $y = 3$ μm, respectively, and their difference is shown in green. The potential difference remains nearly constant across the central part of the sample, away from the edges, corresponding to the Hall voltage that would be measured experimentally.

To understand how this Hall voltage varies with the conductivity of the domain regions, we performed additional simulations while varying the conductivity ratio between the domains and the domain boundaries. **Figure S15b** shows the resulting potential differences for three cases: a ratio of 0.1 (brown), 0.01 (lavender grey), and $10^{-5}$ (green, same as in **Figure S15a**). The black dashed line represents the result from a homogeneous geometry with no domain structure. As the domains become more insulating, the calculated Hall voltage converges to that of the homogeneous case. This suggests that when the domains are sufficiently resistive, the inhomogeneous geometry has minimal influence on the measured Hall voltage. In such cases, the Hall voltage continues to accurately reflect the carrier density as in a uniform system, following the relation $n = B/(eR_{xy})$.

# V. Discussion on the aperiodic moiré structure as an effective disorder

The hTG system naturally forms domain regions composed of commensurate moiré sites and domain boundaries with aperiodic structures through moiré relaxation. This structural contrast is clearly illustrated in the simulation result of **Figure S16a**, which displays both periodic and aperiodic arrangements of moiré sites. In this panel, the left and right regions exhibit periodic arrangements, where the red and blue dots—representing the AA sites of two distinct moiré patterns—form a honeycomb-like structure. In contrast, around the $x \approx -50$ nm region, these moiré sites become deformed, deviating from the periodic order.

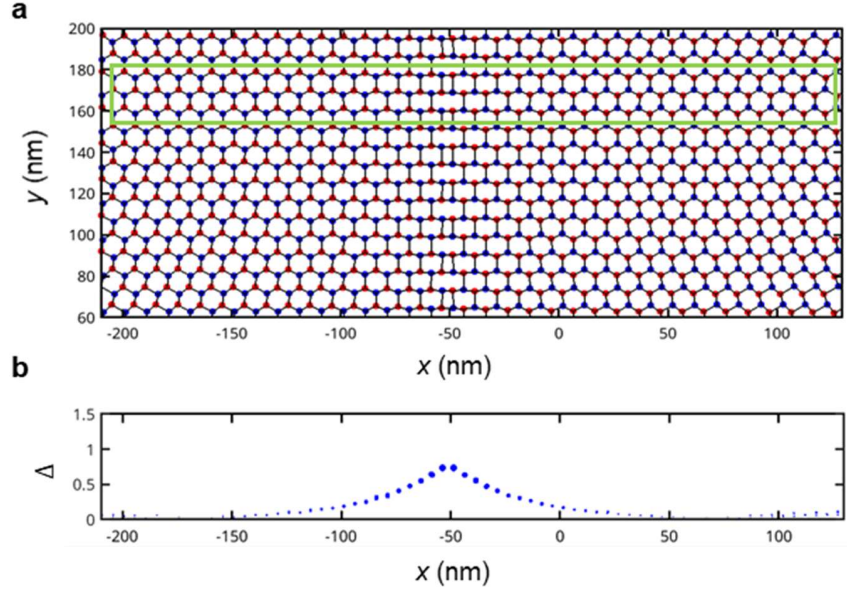

**Figure S16**
**a**, Moiré relaxation simulation results for hTG with a twist angle of 1.4°. The red and blue dots indicate the AA sites of two distinct moiré potentials: one originating from the top and middle layers, and the other from the middle and bottom layers. For visual clarity, black solid lines are drawn between adjacent dots of different colours. Around the $x = -50$ nm region, a domain boundary emerges where the moiré potential sites become aperiodic. **b**, Lattice deformation factor plotted along the x-axis. Within the commensurate domain regions, the factor remains close to zero, whereas it increases sharply near the domain boundary.

Within the domain regions, the periodic moiré sites act as a long-range potential, renormalising the electronic states in the low-energy band. In this context, the aperiodic moiré sites in the domain boundaries can be regarded as an additional perturbation to the periodic moiré potential, which in turn implies that this aperiodicity may act as an *effective* disorder. To characterise the degree of this effective disorder, we introduce the lattice deformation factor, $\Delta$, defined as:

$$\Delta = |\Sigma_{i=1,2,3}\, a_i| / \sqrt{(\Sigma_{i=1,2,3}|a_i|^2)}$$

For each moiré site, we compute the sum of displacement vectors to its three nearest neighbours. $\Delta$ is then obtained by normalising this sum by the square root of the sum of squared vector magnitudes. This value is zero when the moiré sites are periodically arranged in a honeycomb structure, but it increases as the structure deviates from this periodicity. **Figure S16b** shows the lattice deformation factor $\Delta$ plotted along the x-axis from our simulation. As expected, $\Delta$ remains close to zero within the domain regions, while it increases significantly near the domain boundary. These results support the idea that the aperiodic structure at the boundaries introduces an effective disorder. Given this, it is plausible that such disorder could lead to strong localisation of electronic states that would otherwise remain extended.